\documentclass{revtex4}
\usepackage{amsfonts}
\usepackage{amssymb}
\usepackage{graphicx}
\newcommand{\ket}[1]{| #1 \rangle}
\newcommand{\bra}[1]{\langle #1 |}
\newcommand{\avg}[1]{\langle #1 \rangle}

\newcommand{\nket}[4]{|#1\;\;#2\;\;#3\;\;#4\rangle}

\begin{document}
\def\CC{{\rm\kern.24em \vrule width.04em height1.46ex depth-.07ex \kern-.30em 
C}}
\author{K. R. Brown$^1$, K. M. Dani$^2$, D. M. Stamper-Kurn$^2$ and K. B. 
Whaley$^1$}
\title{Deterministic Optical Fock State Generation}
\affiliation{Departments of Chemistry$^1$ and Physics$^2$, University of 
California, Berkeley, California, 94720}
\date{\today}

\begin{abstract}
We present a scheme for the deterministic generation of $N$-photon Fock states 
from 
$N$ three-level atoms in a high-finesse optical cavity. The method applies an 
external 
laser pulse
that generates an $N$-photon output state while adiabatically keeping the 
atom-cavity 
system within a subspace of optically dark states.  We present 
analytical estimates of the error due to amplitude leakage from these dark 
states for 
general $N$, 
and compare it with explicit results of numerical simulations for $N \leq 5$.
The method is shown to provide a robust source of $N$-photon states under a 
variety of
experimental conditions and is suitable for experimental implementation using a 
cloud of 
cold atoms magnetically trapped in a cavity.
The resulting $N$-photon states have potential applications in 
fundamental studies of non-classical states and in quantum information 
processing.

\end{abstract}
\maketitle

\section{Introduction}
\label{sec:intro}

The generation of non-classical states of light has been central
to the confirmation and elucidation of the quantum theory of
radiation.  Today, such work takes on an added importance as part
of the advancing field of deliberate quantum state engineering,
motivated in part by applications in quantum communication and computation.  For
instance,  deterministically generated single-mode single-photon
states could greatly advance the efficiency and security of
quantum cryptography.\cite{qcryptrev} They are also a
crucial resource for implementing a quantum computing scheme
using linear optics, single-photon states, and photodetection
\cite{knil01linear}.  These immediate potential applications have
spurred the development of devices which can produce single photons
on demand, such as solid state devices which use the Coulomb
interaction between strongly confined electrons to produce single
exciton states which then decay optically
\cite{kim99single,mich00single,yuan01},  or devices in which the
fluorescence from single, isolated and optically-pumped molecules is
collected \cite{brun99single,loun00single}. The stream of pulses
produced from each of these devices has been shown to contain either zero
or one, but rarely more than one, photon per pulse, thus
differing radically  from a classical coherent-state pulse which would contain
a Poisson distribution of photon number. Furthermore, while these recent 
devices all
produce fluorescence from a single optical emitter on-demand, none
outputs this fluorescence into a practical single mode of the
optical field.

It has been shown recently that the effects of cavity quantum electrodynamics
(CQED) can in principle be used to overcome this limitation and produce
single-mode single photons on demand
\cite{law97single,kuhn99single}. In such a scheme, a single
three-level atom is induced to fluoresce with high probability
into a single resonant mode of a high-finesse optical cavity.
Experimental evidence for such cavity-induced Raman transitions
has been obtained \cite{henn00raman}, 
and a variant of this scheme has very recently been used to generate a
sequence of single photons \cite{rempexp}.  
By creating a highly controlled, single-quantum-level
interaction of atoms and light, CQED can be used quite generally
to produce non-classical states of the electromagnetic field in a
single cavity mode.
  
In this paper, we analyze a particular
desired function of a CQED device that goes beyond the production of single-
photon states, namely, 
the 
deterministic production of a Fock
state of the electromagnetic field containing an exact number of
photons ($N$).  Such non-classical states are of interest for
fundamental tests and applications of the theories of quantum
optics (such as quantum state tomography, as performed recently on
the single-photon state \cite{lvov01fock}), as a resource for
Heisenberg-limited quantum measurements made possible by the
production of two orthogonal $N$-photon states
\cite{wine92squeeze,bouy97}, and quite generally as a starting point for
the controlled engineering of more complex quantum states.
Starting with a precisely counted number of $N$ atoms trapped in
the confines of a high-finesse optical cavity, we consider a
scheme in which a classical pump field
is ramped up to induce deterministic Raman emission into a single cavity mode by
each of the trapped atoms, resulting in an optical field of exactly $N$ photons 
that are emitted from the cavity in a single pulse.  
The atom-cavity system is constrained to remain within a subspace of 
optically dark $N$-atom states, resulting in a high fidelity of production.  The 
present scheme provides a generalization of that proposed for the production of 
single photons from single atoms \cite{law97single,kuhn99single,rempexp} and 
indicates a 
systematic route to generation of other non-classical states.

Several other
theoretical and experimental works have discussed the use of
high-finesse cavities for the quantum engineering of mesoscopic
non-classical optical states.  
The possibility of producing 
both Fock states and arbitrary coherent superpositions of these inside a cavity 
by
exploiting adiabatic transfer of atomic ground-state Zeeman coherences in single 
atoms
was already explored in Ref.~\cite{parkins93}. 
A scheme for producing an arbitrary quantum optical state using a
single two-level atom in a high-finesse cavity has also been presented by
Law and Eberly, based on the arbitrary real-time control of a
classical pump field and the coupling to a cavity field
\cite{law96arb}. 
Experimental evidence for Fock
states of a microwave cavity field has recently been obtained as a
dynamical equilibrium for a stream of Rydberg
atoms passing through a micromaser \cite{varc00}.  Another
approach toward the construction of a Fock state was proposed in
Ref.\ \cite{domo98fock} in which a Rydberg atom with a
Stark-tunable level splitting is used to transfer photons
one-by-one from a classically-populated cavity field to another
initially empty cavity field.   Initial experimental steps towards this goal
have been recently demonstrated \cite{bert02two}.  

In contrast to the schemes of Refs.\
\cite{varc00,domo98fock,law96arb} which require delicate temporal
control of the atom-cavity coupling, 
our present scheme yields the desired 
$N$-photon state for quite arbitrary temporal profiles of the
classical pump field. This is achieved through a rapid adiabatic
passage which transfers the initial ``non-classical'' state of
$N$ atoms to the non-classical state of $N$ photons within a
short period of time.  Thus, the quantum nature of the
photon field is already guaranteed by the initial state of the
atoms trapped in an initially empty cavity.  
The initial conditions can be achieved by lowering a cloud of cold atoms into 
the cavity,
and pulse control is then entirely contained in the time dependence of the 
ramping 
field.  This avoids
the need to make use of atomic motion in controlling the coupling to the cavity 
field,
as was required in Ref. \cite{parkins93}, and is one of the key elements 
allowing
{\it deterministic} production of $N$-photon states to be achieved.
The number of atoms in the cavity can be determined by detection of excited 
state atoms at the single-atom level, as a result of the cavity-mediated shifts 
of atomic levels.
Such single-atom detection has already been demonstrated \cite{hood98}, and 
can be readily extended to larger numbers
of atoms.  Consequently, the present scheme opens the way to
deterministic generation of more complicated quantum states of light by
first producing non-classical states of trapped atoms (such as
spin-squeezed states produced through interatomic
interactions \cite{sore01ent,ragh01dicke} or by measurement
\cite{kuzm00}), and then transferring that state onto the optical
field using CQED.

The remainder of the paper is constructed as follows.  A brief review of the 
deterministic single-photon generation schemes of 
Refs.\ \cite{law97single,kuhn99single} is given in Section~\ref{sec:single}, 
which
establishes some common concepts with the present work. A discussion of the 
$N$-atom/cavity system in a single-mode 
external field follows in Section~\ref{sec:Nphoton}.  We demonstrate here the 
existence 
of a 
family of optically-dark coupled $N$-atom/cavity states and show generally how 
adiabatic
ramping of an external field can be used to generate $N$-photon emission from 
the cavity.
Detailed analysis of the energy spectrum of the
closed $N$-atom/cavity system as a function of the ramp time provides estimates 
of the
populations in the bright states and also leads to analytic estimates of the 
energy gap
required for limiting adiabatic state transfers. In Section~\ref{sec:open} we 
then present analysis of the cavity decay responsible for the $N$-photon 
emission, 
treating in detail the effects of spontaneous emission and non-adiabaticity on 
the output 
states.  We obtain
analytic estimates of error rates deriving from these contributions that scale 
linearly 
in the total number of atoms, 
resulting in a constant relative error in the output $N$-photon states and 
guaranteeing production of an $N$-photon state with high fidelity.
In Section~\ref{sec:quantumtraj} we present numerical simulations of the
open system using the quantum jump approach. These numerical
calculations are used to explore the sensitivity of the scheme to critical 
experimental 
parameters, as well as to 
explore the limits of our analytical estimates of the error bounds.  Finally,
in Section~\ref{sec:conclusions} we summarize and indicate directions for 
further 
work and for experimental implementation.

\section{Atomic states and Deterministic Single Photon Generation}
\label{sec:single}

\begin{figure}
\includegraphics[width=0.75\hsize]{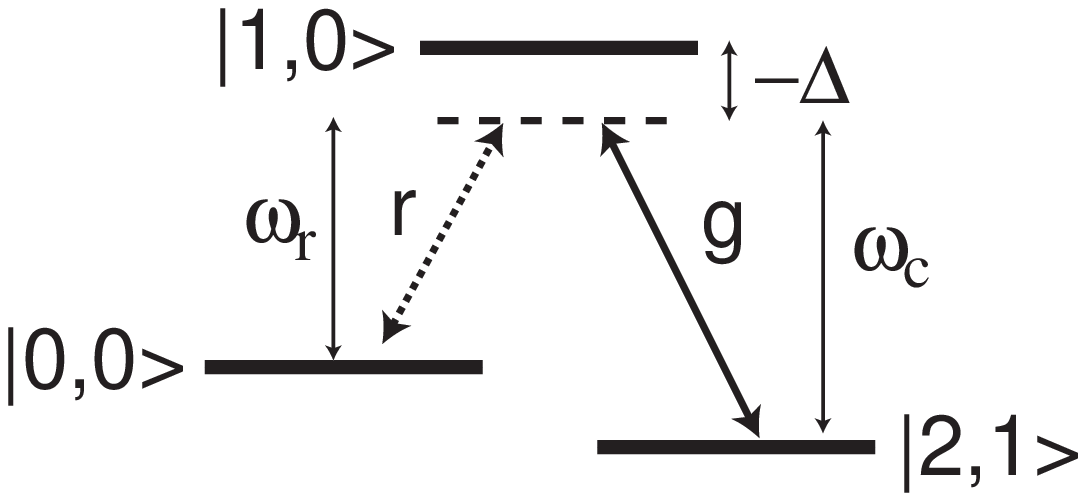}
\caption{Pictorial representation of the Hamiltonian for a single three-level 
atom in a single mode cavity. The atom is driven by an external driving field of 
frequency $\omega_r$ and coupling strength, $r$. The atom is also coupled to a 
cavity mode of frequency $\omega_c$ and coupling strength $g$.  Both the cavity 
and the driving fields are detuned from the atomic transition resonance by a 
common frequency $\Delta$. The atom/cavity states are denoted here as 
$\ket{i,l}$ where $i=0,1,2$ are the three atomic levels and $l$ is the number of 
cavity photons.}
\label{fig:three_level_atom}
\end{figure}

Our $N$-photon generation scheme uses the same internal atomic structure used in
the single-photon proposals of Refs.\ \cite{law97single,kuhn99single}, namely
atoms having three internal levels, labeled
$|0\rangle$, $|1\rangle$ and $|2\rangle$, arranged in a $\Lambda$
configuration such that states $|0\rangle$ and $|2\rangle$ are
non-radiating atomic states 
while $|1\rangle$ is an excited state
connected to states $|0\rangle$ and $|2\rangle$ by allowed
transitions (Figure \ref{fig:three_level_atom}).  
Levels $|0\rangle$ and $|1\rangle$ are typically hyperfine levels of the 
electronic 
ground state. We assume that
the allowed transitions to state $|1\rangle$ can be addressed
selectively. This can be achieved, {\it e.g.}, as a result of polarization 
selection rules, or merely due to a large
energy difference between states $|0\rangle$ and $|2\rangle$.  The three-level 
atoms are 
located in a high-finesse optical
cavity which supports a resonant mode having vacuum Rabi frequency
$g$ ({\it i.e.}, the Rabi frequency due to the presence of a single
photon in the cavity mode) that couples states $|1\rangle$ and
$|2\rangle$.  The cavity mode has frequency $\omega_c$ which can
in general be detuned from the atom resonance $\omega_{1 2}$ by 
$\Delta = \omega_c - \omega_{1 2}$ (see Figure \ref{fig:three_level_atom}).

In the schemes of Refs.\ \cite{law97single} and \cite{kuhn99single},
single-photon generation is accomplished by exposing
a single atom in internal state $|0\rangle$ to a classical
laser field of frequency $\omega_r$ and Rabi frequency $r(t)$,
which is controlled dynamically.  The laser frequency, $\omega_r$, is chosen to
be resonant with the cavity-mediated Raman transition between
states $|0\rangle$ and $|2\rangle$, {\it i.e.},\ $\omega_r - \omega_c =
\omega_{2 0}$ or $\omega_r - \omega_{1 0} = \Delta$.  This laser
connects the states $|0\rangle$ and $|2\rangle$ through a
cavity-mediated Raman transition and induces the fluorescence of
a cavity photon by the atom.
The photon exits the cavity into a single
cavity-output mode, and thereby generates the desired single-photon state.  
Refs.\ \cite{law97single} and \cite{kuhn99single} showed that under suitable 
conditions 
on the external pulse field, the single photon can be emitted deterministically.

To understand the operation of such a deterministic device,
it is helpful to first consider the atom/cavity system as a
closed quantum system, {\it i.e.}, we ignore the decay of cavity
photons to cavity-output modes that actually produces
the desired Fock state outside the cavity, as well as the possible
spontaneous decay from the excited state $|1\rangle$ to modes
other than the cavity mode.  Spontaneous emission lowers the
fidelity of deterministic photon generation, and clearly needs to be avoided or 
at least minimized.  We write the basis states of 
the single atom/cavity system as 
$| 0,0 \rangle$, $|1,0 \rangle$ and $|2,1 \rangle$, where the first index 
refers to the atomic state and the second index gives
the number of photons in the cavity, {\it i.e.}, the cavity field is implicitly
assumed to be quantized. The interaction Hamiltonian
of this closed system is given by 
\begin{equation}
\mathbf{H_0}= \left[
\begin{array}{ccc}
0 & r & 0 \\

r & -\Delta & g \\
0 & g & 0
\end{array}
\right]
\label{eq:Hint}
\end{equation}
where $r(t) \sim \mu_{01}E(t)/2$ is the time-dependent coupling to the external 
(classical) laser pump field, $E(t)$.  Unless essential for the analysis, we 
shall 
omit
the explicit time dependence of $r$ to streamline the notation.
As discussed by Kuhn \emph{et al.}\ \cite{kuhn99single}, the
dynamics of this system are governed by the presence of a
null-valued eigenstate $|\Psi_0\rangle$, which is a ``dark state'' containing no 
population in the excited atomic state $|1\rangle$, and which
is therefore immune to spontaneous decay.  This dark
state exists for all values of $r(t)$ and is given by
\begin{equation}
| \Psi_0 \rangle=\frac 1{\sqrt{r^2+g^2}} \left( g | 0 \rangle - r | 2
\rangle \right).
\label{eq:dark_single}
\end{equation}
It is the presence of this dark state that allows the high-fidelity
generation of a single photon in response to a suitable choice of $r(t)$.  
The single-atom/cavity system starts
initially in the state $|0,0\rangle$, which is the dark state for
the initial condition $r(0) = 0$. During a sufficiently slow ramp
of $r(t)$ ({\it i.e.}, a broad pulse of the classical pump laser), the
atom-cavity system can adiabatically follow the dark state, 
Eq.~(\ref{eq:dark_single}).  
For
sufficiently large values of $r(t)$ ($r \gg g$), $|\Psi_0\rangle
\sim |2,1\rangle$, {\it i.e.}, a single cavity photon is produced
with a high degree of certainty.
This cavity photon then rapidly decays from the
cavity, resulting in a deterministic single-photon source that acts within a 
time interval
specified by the period of the external pump.  Law and
Kimble \cite{law97single} and Kuhn \emph{et al.}\
\cite{kuhn99single} have presented numerical calculations to assess
deviations from this ideal behaviour caused by spontaneous emission, and have
explored the
extent to which the photon emission probability can be controlled by modifying 
the
trigger pulse. Recent experiments by Kuhn, Hennrich, and Rempe have produced 
single photons by this method \cite{rempexp}.

\section{Deterministic $N$-Photon Generation: closed system analysis}
\label{sec:Nphoton}

For the generation of arbitrary Fock states
of the electromagnetic field, {\it i.e.}, with arbitrary large photon number 
$N$,
we now consider $N$ such three-level atoms
confined within the optical cavity. We assume that the atoms are 
indistinguishable in all respects. This 
provides an 
important experimental simplification relative to other CQED schemes in which 
the atoms are required to be individually addressable ~\cite{peli95cqed}.
Each atom interacts individually with the laser field and the
cavity mode, just as for the single atom case.  
We make the simplifying
assumption that the atoms do not interact directly with one
another.  However, they do experience a second order
interaction via the cavity mode.  The
cavity coupling $g$ and the classical pump Rabi frequency $r$
are taken to be identical for each of the $N$ atoms, consistent with their 
indistinguishability \cite{gfootnote}. Thus the
$N$-atom/cavity interaction Hamiltonian is given, for the closed system, by
\begin{equation}
H= \sum_{i=1}^N H_i
\end{equation}
where $H_i$ describes the
atom/field coupling, Eq.~(\ref{eq:Hint}), for the $i{\mbox{th}}$ atom. From now 
on, we 
shall use this in the operator form
\begin{equation}
H_i=-\Delta | 1 \rangle_i \langle 1 |_i +r(t)\left(| 1
\rangle_i\langle 0 |_i+| 0 \rangle_i\langle 1 |_i\right) +g \left(a| 1
\rangle_i\langle 2 |_i +a^{\dagger}| 2 \rangle_i\langle 1 |_i\right),
\label{eq:Hint_op}
\end{equation}
where the operators $a$ and $a\dagger$ are the annihilation and creation 
operators
for the quantized cavity mode.
Since this Hamiltonian is symmetric under the exchange of any two
atoms, a symmetric initial state remains symmetric as it evolves.
We may thus reduce our effective Hilbert state to consider only the states 
that
are completely symmetric with respect to atomic interchange.  We
may thus use a number
representation of the atomic state, namely
$|n_{0}, n_{1}, n_{2}, l\rangle$, where $n_{i}$ gives the
number of atoms in state $|i\rangle $ ($i=0, 1, 2$), and $l$ 
the number of photons in the cavity.  We allow
arbitrary values for $n_0$, $n_1$, $n_2$ and $l$, and employ the
Bose creation and annihilation operators $b_i$ and $b_i^\dagger$
for the atomic states $|i\rangle$. 
We can then rewrite the interaction Hamiltonian for the closed system as
\begin{equation}
H=-\Delta b_1^{\dagger}b_1 + r(t) \left( b_1^{\dagger}b_0 +
b_0^{\dagger}b_1 \right) + g \left( b_1^{\dagger}b_2 a +
a^{\dagger}b_2^{\dagger}b_1 \right). 
\label{eq:H0}
\end{equation}

\begin{figure}
\includegraphics[width=0.75\hsize]{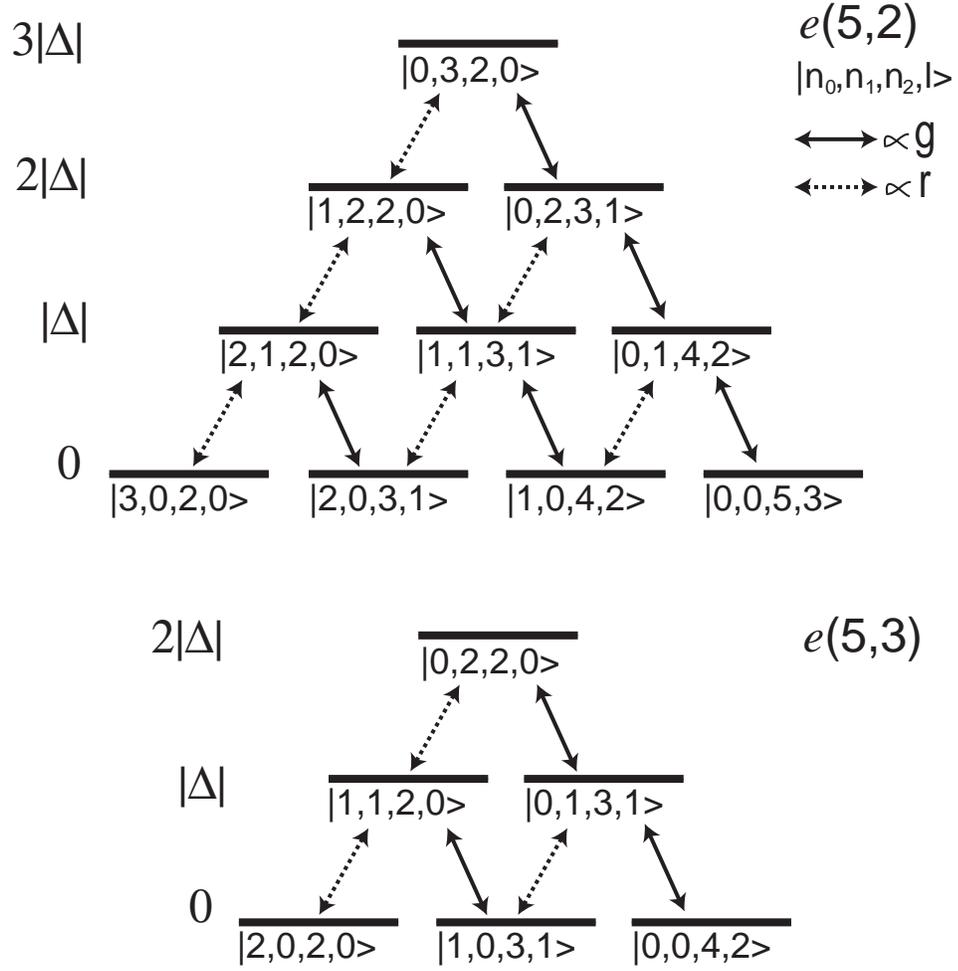}
\caption{Pictorial representation of the Hamiltonian for 5 atoms in the {\it 
e}(5,3) and {\it e}(5,2) manifolds written in the number basis representation 
and assuming red detuning, $\Delta  < 0$.  The $e(N,k)$ manifold is composed of
the eigenstates with simultaneous eigenvalues $N = n_0 + n_1 + n_2$ 
and
$k = n_2 - l$ where $n_i$ is the number of atoms in atomic state $i$ and $l$ is 
the number of photons in the cavity. The transition strength between levels is 
proportional to either the driving field, $r$, or the cavity coupling, $g$.  The 
energy of the states in the absence of all couplings is given by -$n_1\Delta$. 
For red detuning, $\Delta <0$, the states with a higher occupation of the exited 
atomic state, $n_1$, will have a greater energy.  The dark state is the 
superposition of states with $n_1=0$ described in Eq. (\ref{eq:dark}). A 
transition from the manifold {\it e}(5,3) to {\it e}(5,2) occurs when a photon 
is emitted from the cavity.  This transition preserves the dark state (see 
text).}
\label{fig:n_atoms}
\end{figure}

This many-body $N$-atom/cavity Hamiltonian conserves the total number of atoms,
represented by the operator $T = b_0^{\dagger}  b_0 +
b_1^{\dagger} b_1+b_2^{\dagger} b_2$, as well as the difference
between the number of atoms in state $|2\rangle$ and the number of cavity 
photons, represented by the operator $D =
b_2^{\dagger} b_2 - a^{\dagger} a$.  When cavity decay is added to this
description, the operator $D$ gains the significance of referring to the number 
of
photons that have escaped the cavity.  Since the operators 
$T$ and $D$ commute,  we define subspaces $e(N,k)$ composed of
the eigenstates with simultaneous eigenvalues $N = n_0 + n_1 + n_2$ 
and
$k = n_2 - l$ of the operators $T$ and $D$, respectively. This is summarized 
schematically in Figure \ref{fig:n_atoms}.

We find that each manifold $e(N,k)$ contains a null-valued
eigenstate $|\psi_k^N\rangle$, given explicitly by 
\begin{equation}
\left| \psi _{k}^{N}(t)\right\rangle
=\frac{1}{Z_{k}}\sum_{j=k}^{N}\frac{ (-r(t)/g)^{j}}{\sqrt{(N-j)! j!
(j-k)!}}|N-j, 0, j, j-k\rangle 
\label{eq:dark}
\end{equation}
where $Z_{k}$ is a normalization constant. 
This state is the analog of the null-valued
dark eigenstate for the single-atom/cavity system~\cite{kuhn99single}. It 
contains
no population in the $|1\rangle$ internal state and is thus an $N$-atom
dark state immune to spontaneous decay from any atom.
Eq.~(\ref{eq:dark}) represents a continuous family of dark states that are 
transformed into each other by the time dependence of $r(t)$.
For nearly all values $\Delta \neq 0$, we can show that the
null-valued eigenspace of $H$ is non-degenerate, {\it i.e.}, each of these dark 
eigenstates
possesses a unique energy.  We will discuss exceptions later 
(Section~\ref{sec:deltadelta}).
One can also show that acting on Eq.~(\ref{eq:dark}) with the cavity 
annihilation
operator $a$, produces the corresponding dark state having one less photon,
{\it i.e.}, any such dark eigenstate $|\psi_k^N\rangle$ decays
to $|\psi_{k+1}^N\rangle$ by cavity emission. Thus cavity decay does not take 
the system out of the family of 
dark
states.  Conversely, the only way to directly couple dark states in 
different
manifolds, is either to spontaneously lose a photon or to add a particle. 

This representation suggests that adiabatic evolution might be used for 
$N$-photon 
generation 
in an analogous manner to that proposed for single-photon generation 
in 
Refs.\ \cite{law97single,kuhn99single}. We illustrate this here with the 
manifold
corresponding to $k=0$.  Initially the pump laser is off, $r=0$, and the 
corresponding initial state has $N$ atoms in the $|0\rangle$ state and no 
photons in the cavity, $|\psi_k^N(0)\rangle=|N,0,0,0\rangle$.  One can then 
imagine slowly ramping the value of the pump laser until the pump laser coupling 
is much larger than the atom coupling to the cavity, $r \gg g$.  At this final 
time, $t_f$, one finds that approximately all $N$ atoms are in the state 
$|2\rangle$ and there are $N$ photons in the cavity, $|\psi_k^N(t_f)\rangle \sim 
\ket{0,0,N,N}$. 

This procedure will generate $N$ photons in a closed cavity. We will show that 
this procedure can be coupled to cavity decay to produce an $N$-photon state. In 
practice, realization of this ideal sequence requires that two key issues be 
adequately 
addressed.  
First, the need for
adiabatic evolution through the family of dark states via couplings to the 
excited atomic
level $|1\rangle$ places constraints on how we vary the strength of the pump 
pulse, $r$, based on the energy spectrum of the $N$-atom/cavity system.
Second, spontaneous emission of photons from the cavity will provide 
perturbations to the adiabatic 
evolution
that may be non-negligible.

We examine these issues in detail in Section IV.  Before this, we first analyze 
the
energy spectrum of the closed $N$-atom/cavity system in the remainder of
this Section. This will allow us to establish the critical 
parameters limiting the adiabatic evolution, that are required in order to 
estimate the 
errors due to non-adiabaticity and spontaneous emission within the open system
approach employed in Section~\ref{sec:open}.  
While the dark state of each $e(N,k)$ manifold has a succinct description, 
Eq.~(\ref{eq:dark}), the general eigenstates for the $N$-atom/cavity system are 
quite 
complicated. However, in the limit of both large and small $r(t)$, the 
eigenstates are
found to have familiar forms that render them susceptible to analytic 
investigation. 

\begin{figure}
\includegraphics[width=0.75\hsize]{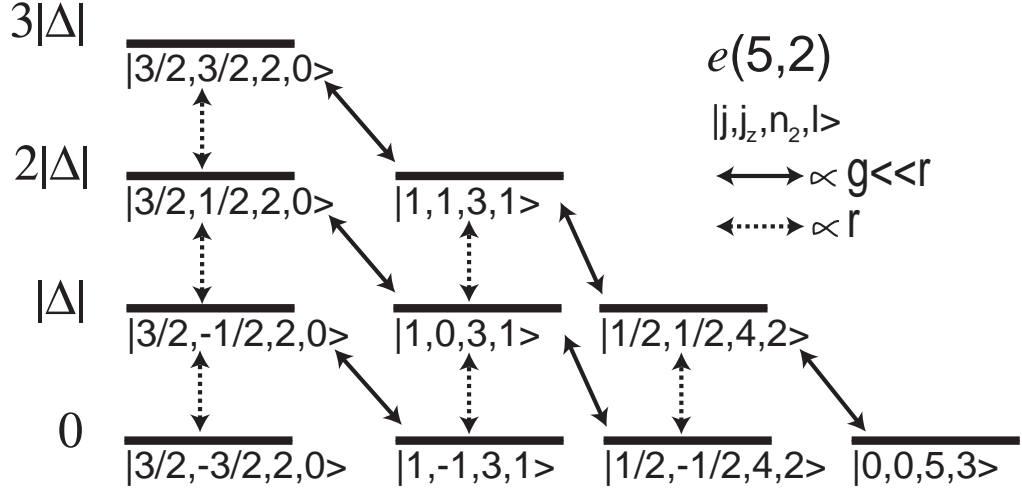}
\caption{Pictorial representation of the Hamiltonian for 5 atoms in the {\it 
e}(5,3) manifold using the Schwinger representation for atomic modes $\ket{0}$ 
and $\ket{1}$ and assuming red detuning, $\Delta  < 0$.  This representation is 
appropriate when the external driving field, $r$, is much larger than the cavity 
coupling, $g$.  The eigenstates to first order are eigenstates of angular 
momentum about an axis defined by the effective magnetic fields $B_z=-\Delta$ 
and $B_x=r$.  The cavity coupling, $g$, acts as a perturbation, mixing states 
differing by a total Schwinger angular momentum of 1.}
\label{fig:ang_mom}
\end{figure}

We examine first the strong pump (large $r$) limit.  Here the eigenstates can be 
interpreted 
in
terms of the familiar angular momentum states.
In the limit where $r$ is large relative to the other parameters ($g$, 
$\Delta$), we can 
neglect the terms in Eq.~(\ref{eq:H0}) that are 
proportional to the cavity coupling parameter $g$.  We then make a 
transformation from 
the atomic modes $b_0$ and $b_1$ to generalized angular momentum operators, $J$, 
using 
the Schwinger 
representation.\cite{schwinger,sakurai} This gives
$J_z=\frac{1}{2}(b_1^\dagger b_1-b_0^\dagger b_0), J_+=b_1^\dagger b_0, J_-
=b_0^\dagger b_1,$ 
and $J_t=\frac{1}{2}(b_1^\dagger b_1+b_0^\dagger b_0)$, where $J_t$ denotes the 
total 
angular momentum, ${\bf J}^2=J_t(J_t+1)$. 
In this limit we find that the system Hamiltonian becomes 
\begin{equation}
H=-\Delta (J_t+J_z)+2r(J_x)=-\Delta J + \Omega ( \hat{\eta}\cdot {\bf J}),
\label{eq:rlarge}
\end{equation}
where $\Omega=\sqrt{4r^2+\Delta^2}$, $\eta_x=2r/\Omega$, $\eta_y=0$, and 
$\eta_z=-\Delta/\Omega$.  The corresponding energy levels are now identical to 
those
of the generalized angular momentum $J$.  Therefore the eigenstates are simply 
given by $\ket{j,m_\eta,n_2,l}_J$ where $j$ is the eigenvalue of total angular 
momentum, 
$m_\eta$ is the angular momentum projection along the axis ${\bf \eta}$, 
$n_2$ is the number of atoms in atomic state $|2\rangle$, and $l$ is the number 
of 
cavity photons as before. 
The cavity coupling term, $ g \left( b_1^{\dagger}b_2 a +
a^{\dagger}b_2^{\dagger}b_1 \right) $, can now be considered to act 
perturbatively 
on Eq.~(\ref{eq:rlarge}), to
mix states differing by $j=\pm\frac{1}{2}$, and to change the value of the 
cavity photon 
number by unity.
Relating these states in the large $r$ limit to our invariants $T$ and $D$, we 
find that 
the $e(N,k)$ manifold is composed of states $\ket{j,m_\eta,n,l}$,
where $0\leq j\leq (N-k)/2$,$n=N-2j$, and $l=N-k-2j$ (see Figure 
\ref{fig:ang_mom}).  
Our dark state at large $r$ is $\ket{0,0,N,(N-k)}_J$. In the number 
representation this
is simply $\ket{n_0=0,n_1=0,n_2=N,l=(N-k)}$ (Eq.(\ref{eq:dark})). We note that 
the loss 
of a photon only reduces the photon number $l$, and conserves $j,m_\eta,$ and 
$n_2$.  The eigenstates in the large $r$ limit are shown 
schematically in
Figure~(\ref{fig:ang_mom}).

We now consider the weak pump (small $r$) limit.  Here we find that the 
eigenstates may also 
be
interpreted in terms of a known set of states, but these are now the less well 
known
eigenstates of the Tavis-Cummings Hamiltonian \cite{tavis}.  We proceed in this 
limit by
starting from the the system Hamiltonian, Eq. (\ref{eq:H0}), at $r(t)=0$. This 
is simply
\begin{equation}
H=-\Delta b_1^{\dagger}b_1 + g \left( b_1^{\dagger}b_2 a +
a^{\dagger}b_2^{\dagger}b_1 \right). 
\label{eq:H_r0}
\end{equation}
We again make a transformation to a Schwinger representation, but this time we 
choose the
transformation to be made between modes $b_2$ and $b_1$.  The generalized
angular momentum operators that are created from these two modes will be denoted 
here
by $F$, {\it i.e.} $F_z=\frac{1}{2}(b_1^\dagger b_1-b_2^\dagger b_2)$, 
$F_+=b_1^\dagger b_2, F_-=b_2^\dagger b_1,$ 
and $F_t=\frac{1}{2}(b_1^\dagger b_1+b_2^\dagger b_2)$, where $F_t$ denotes the 
corresponding total angular momentum, ${\bf F}^2=F_t(F_t+1)$. 
Eq.~(\ref{eq:H_r0}) then becomes
\begin{equation}
H=-\Delta(F_t + F_z) + g(F_+ a + F_- a^\dagger),
\label{eq:H_tavis}
\end{equation} 
which is recognized to be the off-resonant Tavis-Cummings 
Hamiltonian \cite{tavis}.  Note that this Hamiltonian conserves the generalized
angular momentum $F_t$.  It also conserves sum of the number of photons, $l$, 
and the 
angular momentum in the z direction, $f_z$.
In the small $r$ limit we can then investigate the effect of finite $r$ using a 
perturbative analysis.  This 
perturbative 
analysis has two consequences for the energetics.
First, the perturbation due to $r$, which is of the form 
$b_1^\dagger b_0 +b_0^\dagger b_1$, will only couple states whose total $F$ 
value 
differs by 1/2.  Second, as a result of this, the resulting change in energy of 
the 
eigenstates of Eq.~(\ref{eq:H0}) is only second order in $r$.  The eigenstates 
of the Tavis-Cummings Hamiltonian are not trivial, but we note that in the limit 
of large $\Delta$ they are approximately eigenstates of $F_z$ with a number of 
photons in the cavity given by $l=f-f_z-k$ where $f$ is the total angular 
momentum and $k$ is an eigenvalue of $D$.  The states of the Tavis-Cummings 
model are described in detail in \cite{tavis} and qualitatively in Appendix A.
Figure \ref{fig:tc_levels} provides a schematic of these states in the small $r$ 
limit.

One useful advantage of these Schwinger angular momentum representations of the 
atomic
states for our analysis is that  
in both of these limits of large and small $r$, {\it i.e.}, whether for a fixed 
$f$ or 
for fixed $j$, 
the corresponding eigenvalues $f_z$ and $j_z$ provide a measure of the 
population in the 
excited state.
This population is given by $b_1^\dagger b_1 = F_t+F_z=J_t+J_z$.  This property 
will be 
used in 
Section \ref{sec:deltadelta} to make estimates of the population in the 
scattering 
state {\it i.e.}, in the atomic state $|1\rangle$, that is susceptible to 
spontaneous
emission, and hence of the errors due to spontaneous decay.

\section{Open System Approach}
\label{sec:open}

The above description of the $N$-atom/cavity system as a closed
quantum system is clearly incomplete, since a proper assessment of
the operation of an $N$-photon generator requires the
consideration of this CQED device as an open quantum system. We
must take into account the two channels by which the $N$-atom-cavity
system interacts with its environment. These are: i) the possibility of
spontaneous decay from atoms in the excited state $|1\rangle$ to
optical modes outside the cavity, determined by the spontaneous
decay rate $\gamma$, and ii) the coupling of cavity photons to
electromagnetic modes outside the optical cavity, characterized
by the cavity decay rate $\kappa$ \cite{kappafootnote}. The latter provides the 
required coupling to transfer an $N$-photon state from the cavity mode to an 
external mode.  Cavity decay thus plays two different roles here.  First, in 
allowing emission of the $N$-photon state, and, second, affecting the dynamics 
inside the cavity as discussed below.   We 
describe the $N$-atom-cavity system as an open system within a quantum 
wavefunction formulation \cite{carmichael}. To characterize its action as an N-
photon generator, we evaluate the cavity flux $2\kappa\langle a^{\dagger}a 
\rangle$. 

In this Section, we use the quantum wave function formulation in a perturbative 
regime to derive analytic estimates 
for the error rates of its action as an $N$-photon generator.  In Section
\ref{sec:quantumtraj} we then make numerical simulations of the full
open quantum system that allow us to ascertain the extent of validity of
these perturbative error estimates.

In this open quantum system analysis, we will show that the structure of the 
closed quantum system, namely the identification of manifolds of states
$e(N,k)$, each of which contains a
dark state that is immune to spontaneous emission and that
connects adiabatically to the initial state, still plays a
critical role.  When $r(t)$ is varied in the open quantum system, the $N$-
atom/cavity system evolves primarily
within the family of accessible dark states $|\psi_k^N\rangle$
($k \in \{0, ... N\}$) from which no spontaneous emission occurs, just as in the 
closed quantum system.
The system can fail to produce the desired $N$-photon output
state \emph{only if photons are actually lost to spontaneous emission}.
Thus a crucial part of assessing the failure rate of the
$N$-photon generator is to quantify the extent of
``non-darkness,'' {\it i.e.}, the probability that the system will evolve
towards a bright state from which spontaneous emission may
indeed occur.  Non-darkness can arise from two
factors: non-adiabatic evolution when the rate of 
change of $r(t)$ is too fast, and the conditional dynamics resulting from
cavity decay which can couple the dark state $|\psi_k^N\rangle$ to a non-dark 
state.

In the quantum trajectory approach, \cite{plenioqj,carmichael,wiseman93} the 
dynamics 
of the $N$-atom-cavity system are characterized by deterministic
non-unitary evolution, interspersed with random 'jumps' determined by
photon losses from the cavity and by spontaneous emission (scattering) from the
excited state. The non-unitary evolution is given by the
conditional Hamiltonian
\begin{equation}
H_{cond}(t) = H(t) - i\kappa a^{\dagger} a-i\gamma
b_1^{\dagger}b_1. \label{hcond}
\end{equation}
The imaginary terms $- i \kappa a^\dagger a$ and $- i \gamma
b_1^\dagger b_1$ describe the back-action on the quantum system
from that accumulates between instances of cavity decay and spontaneous emission
jumps, respectively.  The corresponding quantum jump operators are given by 
$\kappa a$ and $\gamma b$.  For a full discussion of the quantum jump approach 
see Ref. \cite{carmichael}.   

For the numerical calculations
presented in Section \ref{QTS} we use the full conditional
Hamiltonian, as required in the quantum trajectory 
formulation.\cite{carmichael,plenioqj,wiseman93}  In order to obtain analytic 
error 
estimates here, we proceed first by assuming that the adiabatic errors are
small.  Therefore, at all times the state of the system is now regarded as 
deviating
only slightly from the dark state.  In this situation we may neglect the
spontaneous emission term, $-i \gamma b_1^\dagger b_1$, treating it implicitly
as a higher order perturbation than the cavity decay term.  The cavity decay 
term $\kappa$ is treated in first order perturbation theory.
Before describing the details of this analysis, 
we note again that the cavity
decay, characterized by the jump operator $\kappa a$, connects
dark states in the manifold $e(N,k)$ to dark states in
a lower manifold $e(N, k+1)$, i.e.,\
\begin{equation}
a\left| \Psi _{k}^{N}\right\rangle =\frac{rZ_{k+1}}{gZ_{k}}\left| \Psi
_{k+1}^{N}\right\rangle .
\label{jump}
\end{equation}
(See Section~\ref{sec:Nphoton} above.)  Thus neither of
the quantum jump operators $\kappa a$ or $\gamma b_1$ will lead to
errors (i.e.\ to non-darkness) in the operation of our CQED
device, as long as the system is maintained within the dark states 
$|\Psi_k^N\rangle$.  Consequently, in order to quantify the failure rate of
the $N$-photon generator under these conditions, one need only consider the 
evolution of the system
under the conditional Hamiltonian $H_{cond}$.  We will now make a 
detailed analysis of the errors, starting with an estimate of the extent of non-
darkness 
introduced by cavity decay, then estimating the adiabatic error in 
following the dark state
as $r(t)$ is varied, and finally estimating the spontaneous emission flux rate 
due to 
population in the
$|1\rangle$ internal state that is introduced by a combination of cavity decay 
and non-adiabaticity.

\subsection{Effect of Cavity Decay on Dark States}
\label{sec:cavity}

We first analyze the errors due to cavity dynamics. For this purpose, we treat 
the
$i \kappa a^\dagger a$ term in Eq.\ (\ref{hcond}) as a perturbation
to the closed-system Hamiltonian, Eq.\ (\ref{eq:H0}).  In particular,
under the realistic scenario when cavity decay is weaker than the cavity 
coupling
($\kappa < g$), we find that the cavity decay term causes the
dark states $|\psi_k^N\rangle$ of the closed-system Hamiltonian
to be modified to the states $|\tilde{\psi}_k^N\rangle$ of the
conditional Hamiltonian, according to
\begin{equation}
\left| \tilde{\psi}_{k}^{0}(t)\right\rangle =\left| \psi
_{k}^{0}(t)\right\rangle -i\kappa \sum_{i\neq 0}|\psi
_{k}^{i}(t)\rangle \frac{\langle \psi _{k}^{i}(t)|a^{\dagger
}a\left| \psi _{k}^{0}\right\rangle }{\omega _{i}}.
\end{equation}
Here we have modified our notation to define the states $|\psi 
_{k}^{i}(t)\rangle $ as the instantaneous
eigenvalues of the closed-system Hamiltonian $H_{cond}$ with 
corresponding energies $\hbar\omega_{i}$.
Using this expression, we find the degree of
``non-darkness'' due to cavity decay, 
$\epsilon_k^{\mbox{\small cav}} = 
1 - |\langle \tilde{\psi}_{k}^{0}\left|\psi_{k}^{0}\right\rangle |^{2}$,
to be equal to
\begin{equation}
\epsilon _{k}^{\mbox{\small cav}} = \kappa ^{2}\sum_{i\neq
0}\frac{|\langle \psi _{k}^{i}(t)|a^{\dagger }a|\psi
_{k}^{0}\rangle |^{2}}{\left| \omega _{i}\right| ^{2}}.
\label{eq:ecavity}
\end{equation}

From this expression we can generate an upper bound on the degree
of non-darkness, 
$\epsilon_{k}^{\mbox{\small cav}}$.  We first note that
\begin{eqnarray}
\sum_{i\neq 0} |\langle \psi _{k}^{i}(t)|a^{\dagger }a|\psi
_{k}^{0}\rangle |^{2}  & = & \sum_{i}|\langle \psi
_{k}^{i}(t)|a^{\dagger }a|\psi _{k}^{0}\rangle |^{2}-|\langle
\psi _{k}^{0}(t)|a^{\dagger }a|\psi _{k}^{0}\rangle
|^{2}  \\
& =  & \mbox{Var}(a^{\dagger }a)_{\psi _{k}^{0}(t)},
\end{eqnarray}
where $\mbox{Var}(a^{\dagger }a)_{\psi _{k}^{0}(t)}$ is the
variance in the cavity photon number in the dark state
$|\psi_k^N(t)\rangle$.  Our upper bound is then obtained by replacing
the Bohr frequencies $\omega_i$ in Eq.\ (\ref{eq:ecavity}) with the
minimal Bohr frequency, and replacing the time-varying variance
$\mbox{Var}(a^{\dagger }a)_{\psi _{k}^{0}(t)}$ with its maximum. This
results in
\begin{equation}
\epsilon _{k}^{\mbox{\small cav}} \leq \kappa^2 \frac{\max
Var(a^{\dagger }a)_{\psi _{k}^{0}(t)}}{\min \left| \omega
_{i}\right| ^{2}}.
\label{cavar}
\end{equation}
The maximum variance of $a^{\dagger }a$ is bounded by $N-k$ (see
Appendix B) while the minimal Bohr frequency depends on $\Delta$, and will be 
discussed detail in Section \ref{sec:deltadelta}. Our
expression for the bound on the extent of non-darkness due to cavity decay thus 
becomes
\begin{equation}
\epsilon_k^{\mbox{\small cav}} \leq \frac{\kappa ^{2}(N-k)}{\min \left| 
\omega_i \right| ^2}\label{cmax}.
\end{equation}

\subsection{Adiabatic Errors}
\label{sec:adiab_err}

We now examine errors that arise due to non-adiabatic
evolution resulting from a non-zero derivative, $\dot{r}(t)$.  
Using the
standard treatment \cite{messiahref} we estimate the population leakage from the 
adiabatic state at a time t, 
$\epsilon _{k}^{\mbox{\small ad}}=|\langle \Phi (t)|\psi_{k}^0\rangle 
|^{2},$ 
to first order for a given $N$ and $k$ as
\begin{equation}
\epsilon_k^{\mbox{\small ad}}=\sum_{i\neq 0}\frac{|\langle \psi
_{k}^{i}(t)|
\dot{\psi}_{k}^{0}(t)\rangle |^2}{|\omega_{i}|^{2}}.
\label{aderr}
\end{equation}
Here $|\dot{\psi }_{k}^{0}(t)\rangle =\frac{d}{dt}|\psi_{k}^{0}(t)\rangle $. 
We then apply the standard upper bound for adiabatic error \cite{messiahref}, 
given by the square of the maximum angular velocity of the state, divided by the 
square of the minimal Bohr frequency, 
\begin{equation}
\epsilon_k^{\mbox{\small ad}}\leq \frac{\max \langle \dot{\psi} 
_{k}^{0}(t)|\dot{\psi}_{k}^{0}(t)\rangle }{\min 
|\omega_{i}|^{2}}.
\label{standard}
\end{equation}

One can bound the maximum angular velocity, $\max \langle \dot{\psi} 
_{k}^{0}(t)|\dot{\psi}_{k}^{0}(t)\rangle$ , to be smaller than $\max 
(\dot{r}/g)^2 \frac{N-k}{k+1}$ (see Appendix C).  This results 
in the following upper bound:

\begin{equation}
\epsilon ^{\mbox{\small ad}}_k \leq \frac{(N-k)\max (\dot{r})^{2}}{(k+1)g^2 \min 
|\omega_{i}|^{2}}.
\label{admax}
\end{equation}  

Eqs. (\ref{admax}) and (\ref{cmax}) provide an upper bound for the
``non-darkness'' in the limit that our perturbative approach is appropriate. 
When the minimal energy separation, $\min |\omega_i|$ is independent of $N$, 
both equations suggest that the ``non-darkness'' scales at worst linearly with 
$N$, in the 
worst case. Furthermore, we note that the maximum ``non-darkness'' decreases 
with increasing $k$.  We now proceed to estimate the effect of spontaneous 
emission, or equivalently, the rate of spontaneous emitted flux. 

\subsection{Spontaneous emission flux rate}
\label{sec:deltadelta}

The spontaneous emission flux rate is equal to the product of the population in 
the spontaneously 
emitting state and twice the decay rate of that state, {\it i.e.}, 
$2\gamma\langle b_1^\dagger b_1\rangle$.  
Unfortunately, we are unable to analytically calculate the average population in 
the excited atomic state $\langle b_1^\dagger b_1\rangle$.
The simplest way to estimate an upper bound on the spontaneous flux rate is then 
to  use the maximum possible value of $\langle b_1^\dagger b_1 \rangle=N-k$, and 
the maximum probability of the system being in a non-dark state, 
$\epsilon_k=\epsilon_k^{\mbox{\small cav}}+\epsilon_k^{\mbox{\small ad}}$. We
can thus obtain an upper bound on the spontaneous emission flux rate as 
$2\gamma\epsilon_k(N-k)$.

Naturally, by detuning from resonance, we expect to minimize the population that 
will leak to population in the excited state as is the case in both off-resonant 
Rayleigh and Raman scattering.  Hence, we expect the spontaneous emission flux 
rate to 
change as a function of the detuning $\Delta$.  The value of $\Delta$ affects  
the spontaneous error rate in two ways.  First, it controls the value of the 
minimal Bohr frequency, and can thereby either reduce or enhance the first order 
population leakage from the dark state.  Second, we know by analogy with the 
three level system (Section~\ref{sec:single} and 
Refs. \cite{ law97single,kuhn99single}) that the ``non-darkness'' should 
contain less excited state character as the absolute magnitude of $\Delta$ 
increases.  Our scheme works independent of the sign of $\Delta$. However, for 
concreteness, $\Delta$ is assumed to be negative (red detuning) throughout the 
rest of the paper. 

Referring to the discussion in Section III, for small values of $\Delta$, {\it 
i.e.},$|\Delta| < r, g$, there exist states 
with energy only $\sim \Delta$ away from the dark state. In both the small and 
large $r$ limits, these states are the $m=0$ states (Section~\ref{sec:Nphoton}).
Hence, for $|\Delta| < g$, the minimal Bohr frequency is equal to $\Delta$.  
Furthermore, the eigenstates are effectively eigenstates of $J_x$ or $F_ x$ and 
therefore have average value $J_z = 0$.  Therefore, whatever states the dark
state falls into due to either adiabatic error or cavity decay, these states 
will have large population in the excited atomic state $|1\rangle $.  So in the 
small $|\Delta|$ limit our upper bound described above is reasonable.   
Consequently, we use the ``non-darkness'' factors calculated in Eqs. 
(\ref{cmax}) and (\ref{admax}) times $2\gamma(N-k)$.  Hence, we find that the 
error rate for small values of detuning 
$|\Delta|$ is always less than
\begin{equation}
\gamma^\prime=\gamma \left( \frac{(N-k)^2\max 
(\dot{r})^{2}}{(k+1)g^2|\Delta|^{2}}+
\frac{\kappa^2(N-k)^2}{|\Delta|^2}\right).
\end{equation}

In contrast, for large values of $|\Delta|$, the states are effectively 
eigenstates of $J_z$ or 
$F_z$ (Section~\ref{sec:Nphoton}).  For these states, the smallest valued 
eigenstate is the $m=-J$ or $m=-F$ state, and the corresponding perturbative 
energy scales in second order as $E_1 = -(k+1)(g^2+r^2)/\Delta$.  However, this 
smallest energy eigenstate has a population in the excited state
that scales at worse as $(k+1)g^2/\Delta^2$.  We believe that it is a reasonable 
assumption to multiply this smallest eigenstate estimate of the excited state 
population by our calculated ``non-darkness'' factor $\epsilon$, in order to 
arrive at a better estimate of the excited atomic state population.
This procedure yields the followng expression for large values of $\Delta$, 
namely an error rate that is independent of $\Delta$:
\begin{equation}
\gamma^\prime=\gamma \left( \frac{(N-k)\max (\dot{r})^{2}}{(k+1)^2 g^{4}}+ 
\frac{\kappa^2(N-k)}{(k+1)g^2}\right).
\label{eq:gg}
\end{equation}

Careful analysis shows that in the large $|\Delta|$ limit, the state of energy 
$E_1
= -(k+1)(g^2+r^2)/\Delta$ is not always the closest state to the dark 
state. Indeed states can be identified for particular settings of $\Delta$ and 
$r$ which are degenerate with the null-valued dark state.  However, these states 
are shown to be characterized by higher $f$ or $j$, and therefore in the 
perturbative limit 
they do not couple directly with the dark state.  Numerical simulations suggest 
that the situation 
is similar for states which cross the null state in the non-perturbative limit.

\subsection{Small number of photons in the cavity}.
\label{sec:small}

Having obtained error estimates due to cavity decay, non-adiabaticity, and 
spontaneous emission, we now consider the dependence on the cavity photon number 
$l$.
In the limit that one very slowly ramps up $r$, the average number of photons in 
the cavity is always much smaller than 1.  We assume that our dark state is 
simply $\ket{(N-k)0k0}+\eta\ket{(N-k-1)0(k+1)1}$.  For this state $Var(a^\dagger 
a)=\eta^2-\eta^4\leq\eta^2=<a^\dagger a>$, where $\eta=\frac{r}{g}\sqrt{\frac{N-
k}{k+1}}$ (Eq.~(\ref{eq:dark})). In the limit of small 
$\eta$, the variance of photons in the cavity equals the number of photons 
in the cavity. Since the ramp is very slow, we expect that the error rate will 
be due to the cavity dynamics and, therefore, proportional to the 
variance, Eq. (\ref{cavar}).  The rate of photon output is proportional to the 
number of photons in the cavity. Hence, in the case of a small number of 
photons $l$ in the cavity, we expect at worse for the error to scale 
proportional to the number of photons output.  This implies a total error that 
scales at worse proportional to the number of atoms $N$, when $\Delta > g$.

One might ask how can we increase $r$ to be of equal magnitude with $g$, and 
still maintain a small number of photons in the cavity.  The reason is that as 
$r$ increases, there is an increasing chance of a photon being emitted.  If one 
varies $r$ slowly enough, the chance of that occurring before $r$ becomes too 
large is quite high.  At this point, we then change manifolds $e(N,k)$.  
Examining the equation for the dark state (Eq.~(\ref{eq:dark})) and looking at 
only the first two terms, one sees that changing manifolds is equivalent to 
reducing the effective value of $r$.  Compared to the $k=0$ manifold, the 
effective $r'$ 
of the $k$ manifold is equal to $\sqrt{\frac{N-k}{N(k+1)}}r$.  As $N$ increases, 
for small values of $k$ one then needs smaller and smaller values of $r$  for 
the approximation to hold. For example, when $k=0$, $\eta=\sqrt{N}r/g$.  
However, as the value of $k$ increases, $r$ can eventually reach a value 
comparable to the product $Ng$ and still imply a small number of photons in the 
cavity. 

To summarize, in the limit of a small number of photons $l$ in the cavity and 
$|\Delta| > g$, we 
expect the ratio of spontaneous emission flux to cavity flux to be either a 
constant, or a decreasing function with $N$. Explicitly, we expect spontaneous 
flux to be smaller than  $\frac{2\gamma\kappa^2}{g^2}\int  Var(a^\dagger a) dt 
\leq \frac{2\gamma\kappa^2}{g^2}\int \eta^2 dt$ (see Eq. (\ref{eq:gg}) and Eq. 
(\ref{cavar})). The ratio of the spontaneous flux, 
$\frac{2\gamma\kappa^2}{g^2}\int \eta^2 dt$, to the cavity flux, $2\kappa\int 
\eta^2 dt$, is then $\frac{\gamma\kappa}{g^2}$ which is larger than the 
fractional spontaneous loss, the spontaneous flux divided by the expected number 
of photons out, $N$.  Therefore, for current state of the art optical cavity 
technology for which $\kappa=0.1g$ and $\gamma=0.05g$ \cite{kimcavpar}, we 
expect that only 1 of every 200 photons will be lost to spontaneous emission.

\section{Quantum Trajectory Simulations}\label{QTS}
\label{sec:quantumtraj}
The quantum trajectory or quantum jump approach allows one to calculate the 
properties of open quantum systems by averaging over individual quantum 
trajectories.~\cite{carmichael,plenioqj,wiseman93} The basic elements of this 
approach 
were described briefly in Section \ref{sec:open}. Technical details of the 
algorithm are given in Ref.~\cite{carmichael}.  The calculations described below 
average over 5000 trajectories per simulation.    
The simulations were run until the cavity flux and spontaneous emission flux 
were less than $10^{-6}$. This results in 2000-5000 steps of length $dt=0.1/g$, 
depending on the specific parameters of the system. 

The basic quantity we calculate is the cavity flux, 2$\kappa\langle a^\dagger 
a\rangle$, which describes the number of photons emitted from the cavity. For 
these simulations, we have chosen as our figure of merit the fractional 
spontaneous loss.  The fractional spontaneous loss is defined as the number of 
photons lost to spontaneous emission, $N_s$, divided by the expected number of 
photons out, $N$.  We calculate $N_s$ as the product of $2\gamma$ and the 
integral of $\langle b_1^\dagger b_1\rangle$. Due to the statistical error 
resulting from the finite number of trajectories, we have estimated a numerical 
uncertainty of $\pm 3 \%$ in $N_s/N$.

\subsection{Production of $N$-photon state}
Using a simple driving field which increases linearly with time, and realistic 
cavity parameters based on the cavity of Kimble \cite{kimcavpar}, 
($\dot{r}=g/30, 
\kappa=g/10, \gamma=g/20, \Delta=-2g$), we find that the deterministic 
production of $N$-photons within a single pulse with small loss is indeed 
possible.  Figure \ref{fig:nflux} shows the 
output pulse as a function of time for $N\leq5$.
We see the expected linear increase in the area of the output pulse, {\it i.e}, 
the cavity flux 2$\kappa\langle a^\dagger a\rangle$, 
with $N$. We then calculated the loss of photons into the spontaneous emitted 
modes, shown in Figure \ref{fig:nerr}.  When the simulation was run in the limit 
of a small number of photons in the cavity and with $\Delta>g$, we expected that 
the total photon loss will grow at worst linearly, as explained above in Section 
\ref{sec:small}. Therefore, the fractional spontaneous loss should either be 
constant or decrease with increasing $N$.  This expectation is confirmed by 
simulations, see Figure \ref{fig:nerr} and Figure \ref{fig:nexp}.   Furthermore, 
the error is smaller than our expected bound of $0.5\%$ fractional spontaneous 
loss, see Section \ref{sec:small}. 

Although, we 
have made calculations with only a small number of $N$ values here, it appears 
that 
our analytic estimates coupled with this numeric evidence for $N \leq 5$ suggest 
that the deterministic production of large $N$ Fock states when $\Delta\sim g$ 
is indeed possible.
The numerical results summarized in Figure 6 show that for $\Delta \geq g$ we 
have minimal fractional spontaneous losses of approximately 0.3
actually decrease with increasing $N$. Since our analytical results suggest an 
upper bound of 0.5
expect to be able to produce a Fock state containing on the order of 100 photons 
before the total integrated loss due to spontaneous flux equals a single photon.

\subsection{Minimizing the Spontaneous Loss}
In the previous subsection, we used current experimental cavity parameters 
\cite{kimcavpar} and, for simplicity, a 
linear driving pulse.  We now show how one can minimize the 
spontaneous loss by tuning various cavity parameters and modifying the
driving field.

\subsubsection{Pulse Shape}

For linear ramping with a small rate of change, one finds that the fractional 
error actually decreases with increasing $N$ (Figure \ref{fig:nerr}). This is in 
line with
our analytical bounds which suggested that at worse the fractional error should 
be constant for small occupation of photons in the cavity 
(Section \ref{sec:small}).  
Figure \ref{fig:rerr} shows that as the ramp speed increases beyond the 
adiabatic regime, the errors increase rapidly.
One can further reduce the error by using a more sophisticated time-dependent 
driving field.  For 
example, the application of a Gaussian pulse, $r(t)=g\exp{-(t-
t_0)^2/(2\tau^2)}$, 
reduces the fractional spontaneous loss by a factor of 2, relative to the 
fractional spontaneous loss resulting from a linear ramp, as illustrated in 
Figure \ref{fig:nexp}.  One expects that appropriate pulse shaping could further 
lead to an order of magnitude increase in the fidelity.  However, the minimal 
fractional spontaneous loss obtained by pulse shaping is limited by errors due 
to cavity dynamics.  In Figure \ref{fig:nerr}, one sees that this limit is 
approached by ramp speeds of $g/100$.

\subsubsection{Effect of Detuning}
The ability to detune from resonance is one of the basic tools of atomic 
physics.  
Therefore, it is important to determine how the fidelity of our operation scales 
with the detuning, 
$\Delta$. In Section \ref{sec:deltadelta}, our analytical predictions suggest 
that as $|\Delta|$ is increased from $\Delta=0$, the initially large fractional 
spontaneous loss should decrease, eventually reaching a constant non-zero value 
for $|\Delta|>g$. For larger $|\Delta|$, our numerical calculations confirm this 
prediction for $N=2-5$ atoms, Figure \ref{fig:nerr}.  

In Figure \ref{fig:nerr}, one sees that for detunings, $|\Delta|$, smaller than 
the cavity coupling constant, $g$, the error decreases as the detuning 
increases.  
One then sees a relatively flat region, followed by an increase in error as 
$|\Delta|$ increases.  This rise is outside of the predictive ability of our 
analytical model and represents a breakdown in 
the first order perturbative methods when the energy separation between the dark 
state and nearest neighbor state becomes smaller than the ramping speed.  The 
main conclusion from our numerical simulations is that a wide region of 
detunings are nevertheless possible.  In terms of efficiency, (number of photons 
out per time/fractional spontaneous loss), our data suggest that a detuning of 
$|\Delta|\sim (1-10) g$ would be optimal.

\subsubsection{The good cavity limit minimizes errors.}

For large values of $|\Delta|$, one can reduce the error by increasing the value 
of $g$.   As predicted analytically in Section \ref{sec:deltadelta}, the total 
error will reduce as $g$ increases.  Furthermore, if one varies $g$ and 
$\dot{r}$ proportionally, one can numerically observe the $1/g^2$ 
dependence of fractional loss predicted by Eq. (\ref{eq:gg}). This is shown in 
Figure \ref{fig:gerr}.

\section{Conclusions}
\label{sec:conclusions}
We have presented a scheme for the deterministic production of $N$-photon Fock 
states from coupling of $N$ three-level atoms in a high-finesse optical cavity 
to an external field.  The method relies on adiabatic passage through a 
continuous family of dark states that is controlled by the ramping up of the 
external field. We have shown that this procedure can reliably output pulses of 
$N$-photons from the cavity.  We have made a detailed analysis of the errors 
involved in the $N$-photon state production. These result when there is a 
breakdown of adiabaticity and from spontaneous emission.  This error analysis 
yields analytical bounds on the errors which are well reproduced by the results 
of numerical simulations for up to $N$=5 photons.  Our estimates for physically 
realistic cavity and atom/field parameters indicate that this scheme may be used 
reliably to generate states with up to $N$=100 photons.

One way in which such a deterministic Fock state generator may be 
constructed in the near future is using a combination of an ultra-cold 
atomic source and a high-finesse optical resonator (with parameters such as 
those used for the calculations in Section \ref{sec:quantumtraj}) such as 
those used in the works of Kimble and colleagues \cite{Kimble:98}.  Modern 
techniques of laser and 
evaporative cooling in magnetic traps can readily produce cold and tightly 
confined atomic gases which can be transported in vacuum from a production 
region to the confines of an optical resonator.  Using ultra-cold alkali 
atoms, the states $|0\rangle$ and $|2\rangle$ can be chosen to be two 
ground state hyperfine levels which are both magnetically trapped, such as 
the $|F=1, m_F = -1\rangle$ and $|F=2, m_F = 1 \rangle$ hyperfine states of 
$^{87}$Rb.  These may be connected by a $\Delta m_F = 2$ Raman transition 
using excited levels on the $D1$ or $D2$ transitions -- while the strength 
of this transition is suppressed at large detunings $|\Delta|$ from the 
atomic resonance, our work shows that only moderate detunings (several $g$) 
are necessary for high-fidelity operation. The initial $N$-atom state may 
be generated by following a weak microwave excitation with 
atom-number-sensitive selection, or, alternately, by real-time observation 
of the atomic number in a given hyperfine level.  This would set proper 
initial conditions for $N$-photon generation.

Several additional comments are in order.  Throughout this work,
we have assumed that the atoms are identically coupled to the
cavity and pump fields.  Similarly, we assume that these atoms
are indistinguishable in their spontaneous emission to optical
modes outside the cavity.  It would be desirable for the N-photon
generator to operate similarly if these restrictions are eased.
Toward this end, it is still possible to define a family of dark
states for atoms which are not identically coupled to the cavity
and pump fields.  As such, our analytical approach to estimating
the photon losses to spontaneous emission could be similarly
extended to such cases.  We believe this would provide similar
performance to the idealized case we have provided herein, i.e.
the deterministic production of Fock states containing as many as
$N=100$ photons should be possible.  However, one would expect
that while different spatial arrangements of $N$ atoms in a cavity
would all produce $N$-photon Fock states, the specific optical mode
occupied by those $N$-photons would differ.  Applications requiring
many orthogonal, simultaneous pulses of $N$-photons (such as
Heisenberg-limited interferometry) would be constrained by these
differences.

We envisage a number of applications for
deterministically-produced $N$-photon states.  As alluded to in the
above discussion, two orthogonal $N$-photon states can be combined
using linear beam-splitters to create an interferometer which is
sensitive to differential phase shifts between the two arms of
the interferometer which are on the order of $1/N$ (the
Heisenberg limit), rather than the typical $1/\sqrt{N}$
sensitivity (the standard quantum limit) obtained with classical
light pulses \cite{hlimit}.  Two orthogonal $N$-photon states can also be used 
to create a highly entangled state using only measurement and linear optics 
\cite{Dowling:02}. Such highly entangled photon states can be used to perform 
precision measurements \cite{freqmet,qgyro,clocksync}. The CQED device described 
in this work could thus
be used for demonstrations of this interferometric method.  While
the low photon numbers deterministically available from this CQED
device (given current state-of-the-art cavity parameters) would
not yield the precision available from the use of much more
intense classical light sources, there may be applications
requiring high precision at low light levels which are enabled by
this Fock-state generator, e.g. coherent control \cite{rice}. Other applications 
in the field of
quantum information, communciation, cryptography, and computation
are also possible, and we intend to investigate such potential
applications in future work.  Finally, a deterministic N-photon
source would find both basic and applied use for the absolute
calibration of optical detectors, particularly those designed to
be sensitive to multiple photons (as opposed to conventional
avalanche photodiode devices).  Similarly, the $N$-photon generator
can be used as the light source for novel multiphoton
spectroscopy, e.g. REMPI \cite{REMPI}, although the optical frequency range 
which can be
produced by a given atom-based system would be quite limited.

\section{Acknowledgements}
The authors' effort was sponsored by the Defense
Advanced Research Projects Agency (DARPA) and Air Force Laboratory, Air
Force Materiel Command, USAF, under agreement number F30602-01-2-0524, the 
Office of Naval Research under Grant No. FDN 00014-01-1-0826, and the NSF.
DMSK acknowledges support by the Sloan Foundation. 
KBW thanks the Miller Institute for Basic Research in Science for a Miller 
Research Professorship for 2002-2003.  The work of KRB was also supported by the 
Fannie and John Hertz Foundation. We would like to thank Neil Shenvi for useful 
discussions.  After this work was completed we learned that similar $N$-atom 
dark states have been proposed in Ref.~\onlinecite{lukin}.

\section{Appendix A}
In this appendix, we present a qualitative picture of the eigenstates of our 
system in the small $r$ limit,
i.e., a description of the eigenstates of the non-resonant Tavis-Cummings 
Hamiltonian, Eq.~(\ref{eq:H_r0}). First, note that the creation/annihilation 
operators 
associated with the cavity were not present in Eq.~(\ref{eq:H_r0}), the Tavis-
Cummings 
Hamiltonian would become again simply a Hamiltonian describing angular momentum 
($F$) 
about some axis ($\hat{z}$).  In fact, for large enough $\Delta$, one sees that 
is 
indeed the case, and we do merely have eigenstates of angular momentum about the 
$z$ axis. 
In 
order to get a qualitative picture of the Tavis-Cummings eigenstates for general
values of the cavity coupling $g$ and finite values of $l$, it is useful to 
treat the 
cavity creation/annihilation operators as operators that act to enhance the 
"effective magnetic field" in the "$\hat{x}$" direction. For example, when one 
has a 
large number 
of photons, $l$, in the cavity a standard approximation is to replace the 
cavity photon 
creation/annihilation operators with $\sqrt{l}$.  Applying this
transformation to Eq.~(\ref{eq:H_tavis}), one finds that the Hamiltonian 
becomes simply $H=\Delta(F_t + F_z) + g\sqrt{l}F_x$, {\it i.e.}, the cavity 
coupling
has been transformed into an effective magnetic field in the $\hat{x}$ 
direction.

The eigenstates of Eq.~(\ref{eq:H_tavis}) are constructed as follows.  First, 
one starts 
with atomic states that are the eigenstates of $F_z$, 
namely, $\ket{f,f_z}=\ket{n_1}=f+f_z,n_2=f-f_{z}$.  We know 
that the Hamiltonian conserves $D$, the difference in the number of atoms in 
state 
$\ket{2}$ and the number of photons, $l$, in the cavity.  In this 
representation, the 
conserved quantity is $-f+f_z+l \equiv -k$.  We can then append a photon 
state that reflects this conservation symmetry to the atomic states of modes
$|1\rangle$ and $|2\rangle$ identified via $F_z$. 
This results in the atom-photon states $\ket{f,f_z,k}_F=\ket{f,f_z}_F\ket{l=-
k+f-f_z}$.  
The only missing component is now an atomic state of mode $|0\rangle$.
We then append the appropriate number state of the atomic mode $\ket{0}$ such 
that 
$N$ is conserved, according to $N=2f+n_0$.  This then specifies the remaining 
index $n_0$.
We have thereby arrived at a set of states 
$\ket{f,f_z,k,N}=\ket{n_0=N-2f,n_1=f+f_z,n_2=f-f_z,l=-k+f-f_z}$, that may now be 
latex npb.used
as a basis for expansion of the exact eigenstates of Eq.~(\ref{eq:H_tavis}).
Since the Hamiltonian Eq.~(\ref{eq:H_tavis}) conserves $f, k$, and $n$, the 
exact eigenstates 
can be formally written as $\ket{f,m,k,n}_F=\sum_{f_z}c_{mf_z}\ket{f,f_z,k,N}$.  
Naturally, determining the coefficients $c_{mk}$ is not trivial \cite{tavis}.   
For our
purposes here, it is not necessary to find explicit solutions for the 
coefficients. We
require only the main features of the energy spectrum, which are available from 
the
identification of the above basis and are described below.
 
It is important to note that in order for the photon number $l$ to be positive, 
we must
have $f-f_z\geq k$.  For fixed $f$ and $k$ values, this can lead to a truncation 
of 
the possible accessible states.  One finds that for positive $k$, the angular 
momentum
projection, $m$, can have only $2(f-k/2)+1$ values, whereas for negative $k$, 
$m$ 
can take
all $2f +1$ values.  This implies that for $k\leq 0$, the eigenstate degeneracy 
for given 
values of $f$, $N$, and $k$, equals the degeneracy one would expect for a total 
angular 
momentum value of $f$ ({\it i.e.}, $2f+1$).  In contrast, for values $k>0$, the 
eigenstate degeneracy equals the degeneracy expected for a total angular 
momentum value
of $f-k/2$, {\it i.e.}, $2(f-k/2)+1$.  Note that in our scheme we always
assume that there are initially no photons in the cavity, i.e., $k\geq 0$.

In summary, we see that one think of the states $\ket{f,m,k,N}_F$ as being 
effectively "angular momentum" states possessing a total angular momentum $F_t$
equal to $f-k/2$ and angular momentum projection of $m$ about the "magnetic 
axis".
Although this analogy is not perfect because of the spread of actual 
eigenstates over
this basis, it does contain the following important feature.  This is, that
for $\Delta$ larger than $g\sqrt{l_{max}}=g\sqrt{2f-k}$, where $l_{max}$ is the 
maximum 
number of photons in the cavity, the eigenstates are eigenstates of $F_z$, with 
eigenvalue $m-k/2$.  So the large $\Delta$ limit is simplified.  
In contrast, for small and medium size values of $\Delta$, the eigenstates are 
superpositions over a wide range of $f_z$ states.

\section{Appendix B} \label{App}
In this Appendix, we derive an upper bound on the variance of cavity photon 
number for the dark state. In order to reach this bound, we need to first 
outline other properties of the dark state. We begin with a few definitions. The 
dark state in the $E(N,k)$ manifold is given by,

\begin{equation} \label{psik}
\ket{\Psi_k} = \frac{1}{Z_k} \sum_{l=0}^{N-k}
    \frac{(-x)^l}{\sqrt{(N-k-l)!(l+k)!l!}} \nket{(N-k-l)}{0}{(l+k)}{l},
\end{equation}
where $x=r/g$ and $Z_k$ is the normalization constant.

Let $\ket{\Phi_k}$ be the state given by $\frac{d}{dx} \ket{\Psi_k}$. Thus,

\begin{equation} \label{phik}
\ket{\Phi_k}=\frac{1}{Y_k} \left(
    \frac{1}{Z_k}\sum_{l=1}^{N-k}
       \frac{l(-x)^{l-1}}{\sqrt{(N-k-l)!(l+k)!l!}}
                    \nket{(N-k-l)}{0}{(l+k)}{l}
    \quad - \frac{Z'_k}{Z_k} \ket{\Psi_k}
                \right) \label{eq:yk}
\end{equation}
where $Y_k$ is the normalization constant.

We further define the state $\ket{\chi_k}$ given by the normalized
state $H \ket{\Phi_k}$. Therefore,

\begin{equation} \label{chik}
\ket{\chi_k}=\frac{1}{W_k} \sum_{l=0}^{N-k-1}
       \frac{(-x)^l}{\sqrt{(N-k-1-l)!(l+k)!l!}}
            \nket{(N-k-1-l)}{0}{(l+k)}{l},
\end{equation}
where $W_k$ is the normalization constant.

We now present some properties relating these normalization constants
to each other and their derivatives. So let $\avg{l}_k$ be the average
number of photons in the cavity for the darkstate in the $k$
manifold. In other words,

\begin{eqnarray} \label{avgl}
\avg{l}_k
&=&\bra{\Psi_k} a^{\dagger} a \ket{\Psi_k} \nonumber \\
&=&\frac{1}{Z_k^2}
    \sum_{l=0}^{N-k} \frac{l x^{2l}}{(N-k-l)!(l+k)!l!}
\end{eqnarray}

\bigskip
{\bf Property 1}
\begin{equation}\label{zdot}
x \frac{Z_k'}{Z_k} = \avg{l}_k
\end{equation}
{\it Proof:} By definition, $Z_k$ is the normalization constant of the
dark state $\ket{\Psi_k}$. Hence,
\begin{eqnarray*}
&{}& Z_k^2 = \sum_{l=0}^{N-k} \frac{x^{2l}}{(N-k-l)!(l+k)!l!} \\
& \therefore & 2 Z'_k Z_k=
        \sum_{l=1}^{N-k}\frac{2lx^{2l-1}}{(N-k-l)!(l+k)!l!} \\
& \therefore & xZ_k'Z_k=
        \sum_{l=0}^{N-k}\frac{lx^{2l}}{(N-k-l)!(l+k)!l!}
\end{eqnarray*}
Dividing both sides by $Z_k^2$ gives us,

\begin{eqnarray*}
\frac{x Z_k'}{Z_k} &=& \frac{1}{Z_k^2}
        \sum_{l=0}^{N-k}\frac{lx^{2l}}{(N-k-l)!(l+k)!l!} \\
&=& \avg{l}_k \qquad \left[ \textrm{From Eq. (\ref{avgl})}\right]
\end{eqnarray*}

\bigskip
{\bf Property 2}
\begin{equation}\label{xzk1zk}
\left( x \frac{Z_{k+1}}{Z_k} \right)^2 = \avg{l}_k
\end{equation}

{\it Proof:} $Z_{k+1}$ is the normalizaton constant for the state
$\ket{\Psi_{k+1}}$. Hence by definition,
\begin{eqnarray*}
&{}& Z_{k+1}^2 =
    \sum_{l=0}^{N-(k+1)}\frac{x^{2l}}{(N-(k+1)-l)!(l+k+1)!l!} \\
& \therefore & \left( x Z_{k+1} \right)^2 =
    \sum_{l=0}^{N-k-1)}\frac{x^{2(l+1)}}{(N-k-1-l)!(l+k+1)!l!} \\
& \therefore & \left( x Z_{k+1} \right)^2 =
        \sum_{l=0}^{N-k-1)}
        \frac{ (l+1)x^{2(l+1)} }{(N-k-(l+1))!(l+1+k)!(l+1)!} \\
& \therefore & \left( x Z_{k+1} \right)^2 =
        \sum_{l=1}^{N-k)} \frac{ (l)x^{2l} }{(N-k-l)!(l+k)!l!}
\end{eqnarray*}

Dividing both sides by $Z_k^2$, we get the result,

\begin{eqnarray*}
\left(\frac{x Z_{k+1}}{Z_k} \right)^2 &=&
    \frac{1}{Z_k^2}
    \sum_{l=0}^{N-k)} \frac{ (l)x^{2l} }{(N-k-l)!(l+k)!l!}
\end{eqnarray*}
From Eq. (\ref{avgl}), we prove the required result.

\bigskip
{\bf Property 3}

If the system is in the dark state $\ket{\Psi_k}$, then
\begin{equation}\label{var}
Var(a^{\dagger}a)= (x Y_k)^2
\end{equation}

{\it Proof:} It is trivial to see that Property 3 is true when
$x=0$. Let $x \neq 0$. Consider the state $\ket{\Phi_k}$ given by
eq. \ref{phik}. Since $x \neq 0$ we can extend the summation to
$l=0$. Thus,

\begin{eqnarray*}
\ket{\Phi_k}
&=&\frac{1}{Y_k} \left(
    \frac{1}{Z_k}\sum_{l=0}^{N-k}
       \frac{l(-x)^{l-1}}{\sqrt{(N-k-l)!(l+k)!l!}}
            \nket{(N-k-l)}{0}{(l+k)}{l}
    \quad - \frac{Z_k'}{Z_k} \ket{\Psi_k}
                \right) \\
&=&\frac{1}{Y_k Z_k} \sum_{l=0}^{l=N-k}
    \left(l-x\frac{Z'_k}{Z_k}\right)
        \frac{(-x)^{l-1}}{\sqrt{(N-k-l)!(l+k)!l!}}
            \nket{(N-k-l)}{0}{(l+k)}{l} \\
&{}& \qquad \qquad \qquad   \mbox{[Using Eq. \ref{psik}]} \\
&=&\frac{1}{Y_k Z_k} \sum_{l=0}^{N-k}
    \left( l-\avg{l}_k \right)
        \frac{(-x)^{l-1}}{\sqrt{(N-k-l)!(l+k)!l!}}
            \nket{(N-k-l)}{0}{(l+k)}{l} \\
&{}& \qquad \qquad \qquad  \mbox{[Using Property 1]} \\
\end{eqnarray*}

$Y_k$ is the normalization constant of $\ket{\Phi_k}$. Thus,
\begin{eqnarray*}
Y_k^2 &=& \frac{1}{Z_k^2}
    \sum_{l=0}^{N-k} \left(l-\avg{l}_k\right)^2
        \frac{x^{2(l-1)}}{(N-k-l)!(l+k)!l!} \\
\therefore (x Y_k)^2 &=& \frac{1}{Z_k^2}
    \sum_{l=0}^{N-k} \left(l-\avg{l}_k\right)^2
        \frac{x^{2l}}{(N-k-l)!(l+k)!l!} \\
\therefore (x Y_k)^2 &=& \avg{l^2}_k - (\avg{l}_k)^2
\end{eqnarray*}

It easy to see that this is merely $Var(a^{\dagger}a)$.
\bigskip

{\bf Property 4}
\begin{equation}\label{l2}
\avg{l^2}_k = \avg{l}_k \left( \avg{l}_{k+1} +1 \right)
\end{equation}
{\it Proof:}
\begin{eqnarray*}
\avg{l^2}_k & = & \frac{1}{Z_k^2} \sum_{l=0}^{N-k}
        \frac{l^2 x^{2l}}{(N-k-l)!(l+k)!l!} \\
& = & \frac{1}{Z_k^2} \sum_{l=1}^{N-k}
        \frac{l x^{2l}}{(N-k-l)!(l+k)!(l-1)!} \\
& = & \frac{1}{Z_k^2} \sum_{l=0}^{N-k-1}
        \frac{(l+1) x^{2(l+1)}}{(N-k-l-1)!(l+1+k)!l!} \\
& = & \frac{x^2 Z^2_{k+1}}{Z_k^2 Z^2_{k+1}} \sum_{l=0}^{N-(k+1)}
        \frac{(l+1) x^{2l}}{(N-(k+1)-l)!(l+k+1)!l!} \\
& = & \frac{x^2 Z^2_{k+1}}{Z_k^2}
    \frac{1}{Z^2_{k+1}} \sum_{l=0}^{N-(k+1)}
        \frac{(l+1) x^{2l}}{(N-(k+1)-l)!(l+k+1)!l!} \\
& = & \frac{x^2 Z^2_{k+1}}{Z_k^2} \avg{l+1}_{k+1} \\
\end{eqnarray*}
Using Property 2 we get the required result.

\bigskip
{\bf Property 5}
\begin{equation}\label{lessk}
\avg{l}_k \geq \avg{l}_{k+1}
\end{equation}
{\it Proof:}
We need to show that
\begin{equation}
\avg{l}_k - \avg{l}_{k+1} \geq 0
\end{equation}
which is equivalent to showing that
\begin{equation}
Z^2_{k}Z^2_{k+1}\avg{l}_k - Z^2_{k}Z^2_{k+1}\avg{l}_{k+1} \geq 0
\end{equation}
since $Z_j$ is positive for all $j$.

\begin{eqnarray*}
Z^2_{k}Z^2_{k+1}\avg{l}_k - Z^2_{k}Z^2_{k+1}\avg{l}_{k+1} & = & 
Z^2_{k+1}\sum_{l=0}^{N-k} \frac{l x^{2l}}{(N-k-l)!(l+k)!l!} - 
Z^2_{k}\sum_{m=0}^{N-k-1} \frac{m x^{2m}}{(N-k-1-m)!(m+k+1)!m!}\\
& = & \sum_{m=0}^{N-k-1} \frac{x^{2m}}{(N-k-1-m)!(m+k+1)!m!}\sum_{l=0}^{N-k} 
\frac{l x^{2l}}{(N-k-l)!(l+k)!l!}\\ &-& \sum_{l=0}^{N-k} \frac{x^{2l}}{(N-k-
l)!(l+k)!l!}\sum_{m=0}^{N-k-1} \frac{m x^{2m}}{(N-k-1-m)!(m+k+1)!m!}\\
& = & \frac{x^{2(N-k)}}{N!(N-k)!}\left( \sum_{m=0}^{N-k-1} \frac{(N-k-m) 
x^{2m}}{(N-k-1-m)!(m+k+1)!m!}\right)\\ &+& \sum_{m=0}^{N-k-1}\sum_{l=0}^{N-k-1} 
\frac{(l-m)x^{2(m+l)}}{(N-k-1-m)!(m+k+1)!m!(N-k-l)!(l+k)!l!}\\
\end{eqnarray*}
We note that
\begin{equation}
\frac{x^{2(N-k)}}{N!(N-k)!}\left( \sum_{m=0}^{N-k-1} \frac{(N-k-m) x^{2m}}{(N-k-
1-m)!(m+k+1)!m!}\right)\geq 0.
\end {equation}
Thus,
\begin{equation}\label{eq:5.1}
Z^2_{k}Z^2_{k+1}\avg{l}_k - Z^2_{k}Z^2_{k+1}\avg{l}_{k+1} \geq \sum_{m=0}^{N-k-
1}\sum_{l=0}^{N-k-1} \frac{(l-m)x^{2(m+l)}}{(N-k-1-m)!(m+k+1)!m!(N-k-
l)!(l+k)!l!}\\.
\end{equation}
In order to evaluate the sum in Eq. ({\ref{eq:5.1}}), we choose two integers 
between 0 and $N-k-1$, $a$ and $b$. We then evaluate the sum of the two terms 
corresponding to $l=a$, $m=b$, and  $l=b$, $m=a$. We see that
\begin{eqnarray*}
\frac{(a-b)x^{2(a+b)}}{(N-k-1-b)!(b+k+1)!b!(N-k-a)!(a+k)!a!}+\frac{(b-
a)x^{2(a+b)}}{(N-k-1-a)!(a+k+1)!a!(N-k-b)!(b+k)!b!}\\  =  \frac{x^(2(a+b))}{(N-
k-b)!(b+k)!b!(N-k-a)!(a+k)!a!}\left( (a-b)\left(\frac{N-k-b}{b+k}-\frac{N-k-
a}{a+k}\right)\right)\\.
\end{eqnarray*}

Since $\frac{x^(2(a+b))}{(N-k-b)!(b+k)!b!(N-k-a)!(a+k)!a!} \geq 0$, we simply 
need to determine whether $(a-b)\left(\frac{N-k-b}{b+k}-\frac{N-k-
a}{a+k}\right)\geq 0$.
If $a>b$, $a-b$ is positive and $\left(\frac{N-k-b}{b+k}-\frac{N-k-
a}{a+k}\right)$ is positive, so the product is therfore also positive.  If 
$b>a$, 
$a-b$ is negative and $\left(\frac{N-k-b}{b+k}-\frac{N-k-a}{a+k}\right)$ is also 
negative, so therefore the product is still positive .  Hence, for all $a$ and 
$b$
\begin{equation}
 \frac{x^{2(a+b)}}{(N-k-b)!(b+k)!b!(N-k-a)!(a+k)!a!}\left( (a-b)\left(\frac{N-k-
b}{b+k}-\frac{N-k-a}{a+k}\right)\right)\geq 0,
\end{equation}
which implies that
\begin{equation}\label{eq:5.2}
\sum_{m=0}^{N-k-1}\sum_{l=0}^{N-k-1} \frac{(l-m)x^{2(m+l)}}{(N-k-1-
m)!(m+k+1)!m!(N-k-l)!(l+k)!l!}\geq 0.
\end{equation}
By Eq. (\ref{eq:5.1}) and Eq. (\ref{eq:5.2}), we have
\begin{equation}
Z^2_{k}Z^2_{k+1}\avg{l}_k - Z^2_{k}Z^2_{k+1}\avg{l}_{k+1} \geq 0
\end{equation}
and thus
\begin{equation}
\langle l \rangle_k \geq \langle l \rangle_{k+1}.
\end{equation}
\bigskip
{\bf Property 6}\\
If the system is in the dark state $\ket{\Psi_k}$,
\begin{equation}
Var(a^\dagger a)_k\leq (N-k).
\end{equation}

{\it Proof:}
\begin{eqnarray*}
Var(a^\dagger a)_k &=& \avg{l^2}_k-(\avg{l}_k)^2\\
& = & \avg{l}_k(\avg{l}_{k+1}+1)-(\avg{l}_k)^2\\
& { } & \qquad \mbox{[Using Property 4]}\\
& \leq & \avg{l}_k\\ 
& { } & \qquad \mbox{[Using Property 5]}\\
& \leq & N-k.
\end{eqnarray*}

\section{Appendix C}\label{appB}
We determine here an upper bound on the angular velocity of the dark state,  
$\max \alpha_k^2=\max \langle \dot{\psi} 
_{k}^{0}(t)|\dot{\psi}_{k}^{0}(t)\rangle$.  We first note that using the chain 
rule the $\max \alpha_k^2$ is equivalent to determining the maximum value of 
$\dot{x}^2 Y_k^2$ where $Y_k$ is the normalization constant defined in equation 
\ref{eq:yk}).  For convenience, we will define $c_l=\frac{(-x)^l}{\sqrt{(N-k-
l)!(l+k)!l!}}$ and $c_l^\prime=\frac{(-x)^l}{\sqrt{(N-k-l)!(l+k)!l!}}$ 
(throughout this Appendix, we will use $f^\prime$ to represent the derivative of 
$f$ with respect to $x$, and $\dot{f}$ to represent the derivative f with 
respect to $t$).

Since the maximum of the product of two functions is always less than or equal 
to the product of the maximums of each function, the first step is to simply 
take the maximum of  $\dot{x}^2$.   This yields

\begin{equation}
\max \dot{x}^2 =  \left(\frac{\max \dot{r}}{g}\right)^2.
\end{equation}

The second step is to determine the maximum value of $Y_k^2$.  Using the 
notation from Appendix B, we first note that $\langle \Psi_k|\Phi_k\rangle 
=\frac{1}{Y_k}\langle \Psi_k|\Psi_k^\prime\rangle=0$, since $|\Psi_k\rangle$ is 
normalized. From Eq. (\ref{eq:yk}), this implies that 
\begin{equation}
\langle \Psi_k|\sum_{l=1}^{N-k} c_l^\prime \nket{(N-k-l)}{0}{(l+k)}{l} 
=Z_k^\prime.
\end{equation} 
Hence, we can write the following expression for $Y_k^2$,
\begin{equation}
Y_k^2=\frac{1}{Z_k^2}\left(\sum_{l=1}^{N-k} (c_l^\prime)^2 -
(Z^\prime_k)^2\right).
\end{equation}
Since $Y_k^2$ must be positive and both $\sum_{l=1}^{N-k} (c_l^\prime)^2$ and 
$(Z^\prime_k)^2$ are positive, we can write the following inequality
\begin{equation}
Y_k^2\leq\frac{\sum_{l=1}^{N-k} (c_l^\prime)^2}{Z_k^2}.
\end{equation}

Note that $Y_k^2$ is an even function of $x$, so the value of $Y_k^2$ at $x=0$ 
must be a local minimum or maximum.  When one takes the derivative of $Y_k$ with 
respect to $x$, one finds only a single zero at the origin.  The identity, $x^2 
Y_k^2=var(a^\dagger a)$, implies that $Y_k^2$ must go to zero as $x$ increases 
in order to maintain a finite variance.  Therefore, the maximum value of $Y_k$ 
occurs at the origin.  Explicitly calculating the limit at the origin yields
\begin{equation}
\max Y_k^2=\frac{N-k}{k+1}.
\end{equation}  
Consequently, we have an expression for the maximum angular velocity
\begin{equation}
\max \alpha_k^2= \frac{(N-k)\max \dot{r}^2}{(k+1)g}.
\end{equation}

\begin{figure}
\includegraphics[width=0.75\hsize]{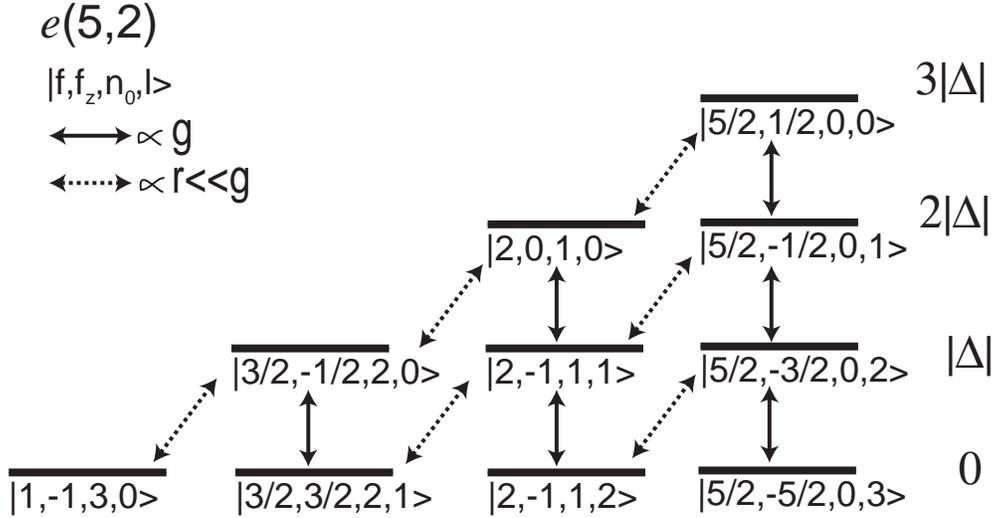}
\caption{Pictorial representation of the Hamiltonian for 5 atoms in the {\it 
e}(5,3) manifold in 
the Tavis-Cummings basis for atomic modes $|1\rangle$ and $\ket{2}$ and assuming 
red detuning, $\Delta<0$. This representation is appropriate when the cavity 
coupling, $g$, is much larger than the external driving field, $r$.  The 
eigenstates of the Tavis-Cummings Hamiltonian are complicated (see text and Ref. 
\cite{tavis}) but preserve total Schwinger angular momentum $f$ obtained from 
modes $|1\rangle$ and $\ket{2}$. The coupling to the external field, $r$, acts 
as a perturbation, mixing states differing by a total Schwinger angular momentum 
of 1.}
\label{fig:tc_levels}
\end{figure}
\begin{figure}
\includegraphics{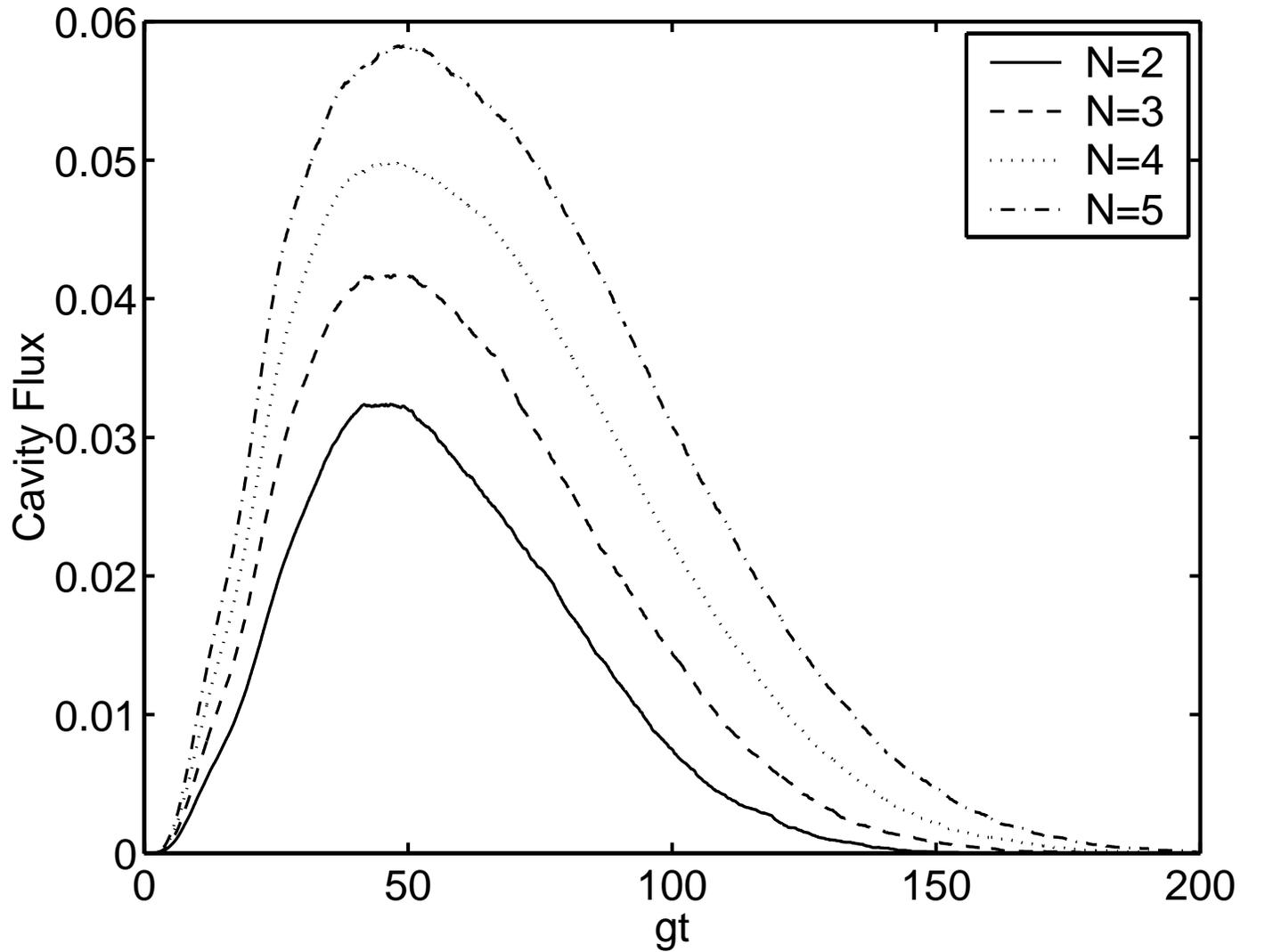}
\caption{The cavity flux, $2\kappa\langle a^\dagger a \rangle$, for $N=2-5$ 
atoms plotted versus dimensionless time, $T=gt$. These simulations were done 
using  a linear ramp $r(T)=RT$ where $R=g/100$.  The other cavity parameters 
were $\gamma=g/20, \kappa=g/10$, and $\Delta=-2g$. The integrated cavity flux, 
$N_s$, increases linearly with $N$ as expected.} 
\label{fig:nflux}
\end{figure}
\begin{figure}
\includegraphics{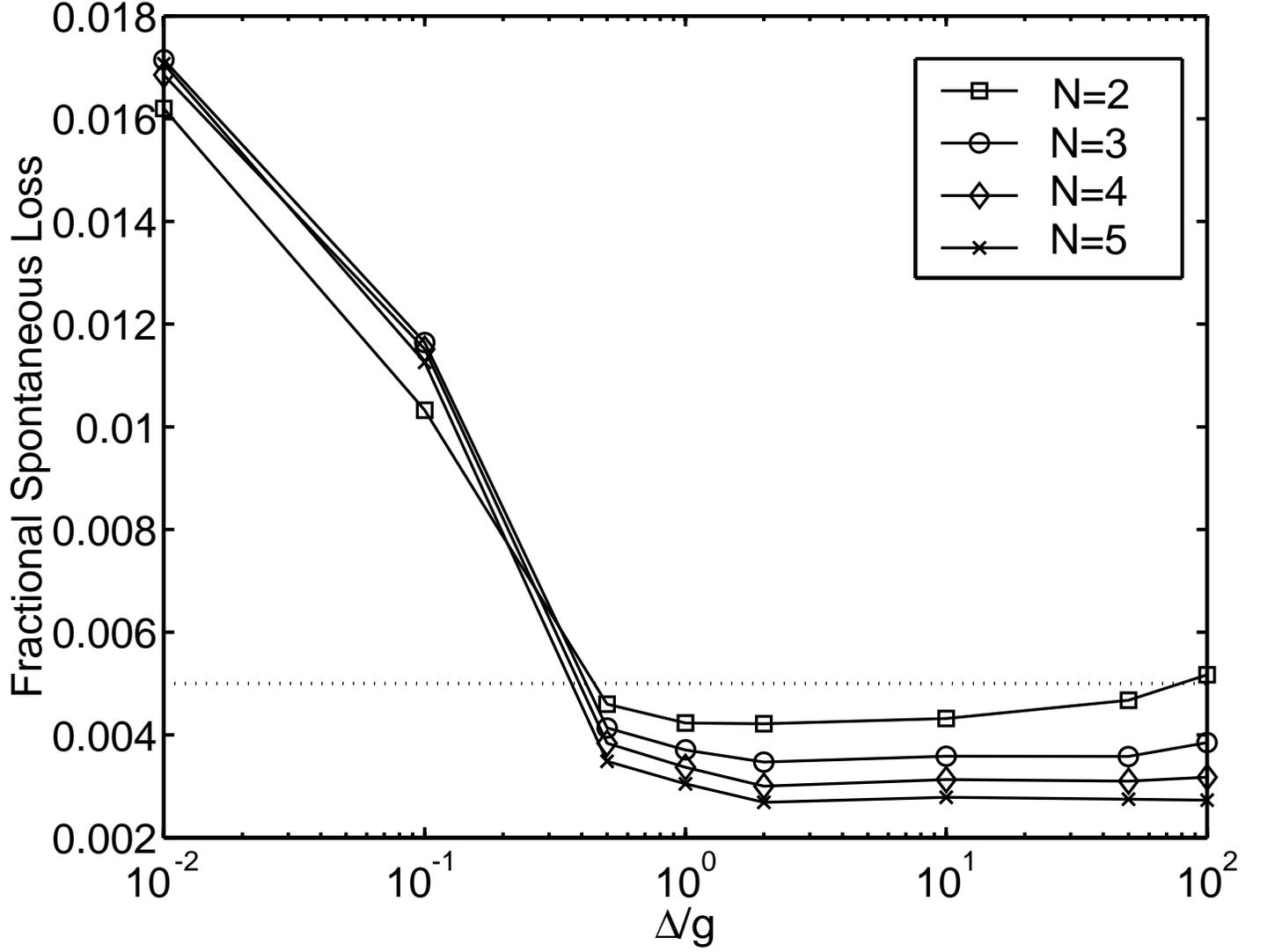}
\caption{ Fractional spontaneous loss for $N=2-5$ as a function of $\Delta$. 
A linear ramp was used with $R=g/100,\gamma=g/20$ and $\kappa=g/10$.  As 
expected one sees that in the large 
$\Delta$ limit that the fractional spontaneous loss scales less than N (see 
text).  Our analytical bounds in the large $|\Delta|$ limit suggest that the 
fractional spontaneous loss should be less than 
$\frac{\gamma\kappa}{g^2}=5\times 10^{-3}$ (dotted line). The discrepancy at 
large $|\Delta|$ represents parameters for which our perturbative approach is 
invalid. For the small $|\Delta|$ limit the increasing fractional spontaneous 
loss is  consistent with leakage to states that have a higher occupation of the 
excited state.}
\label{fig:nerr}
\end{figure}
\begin{figure}
\includegraphics{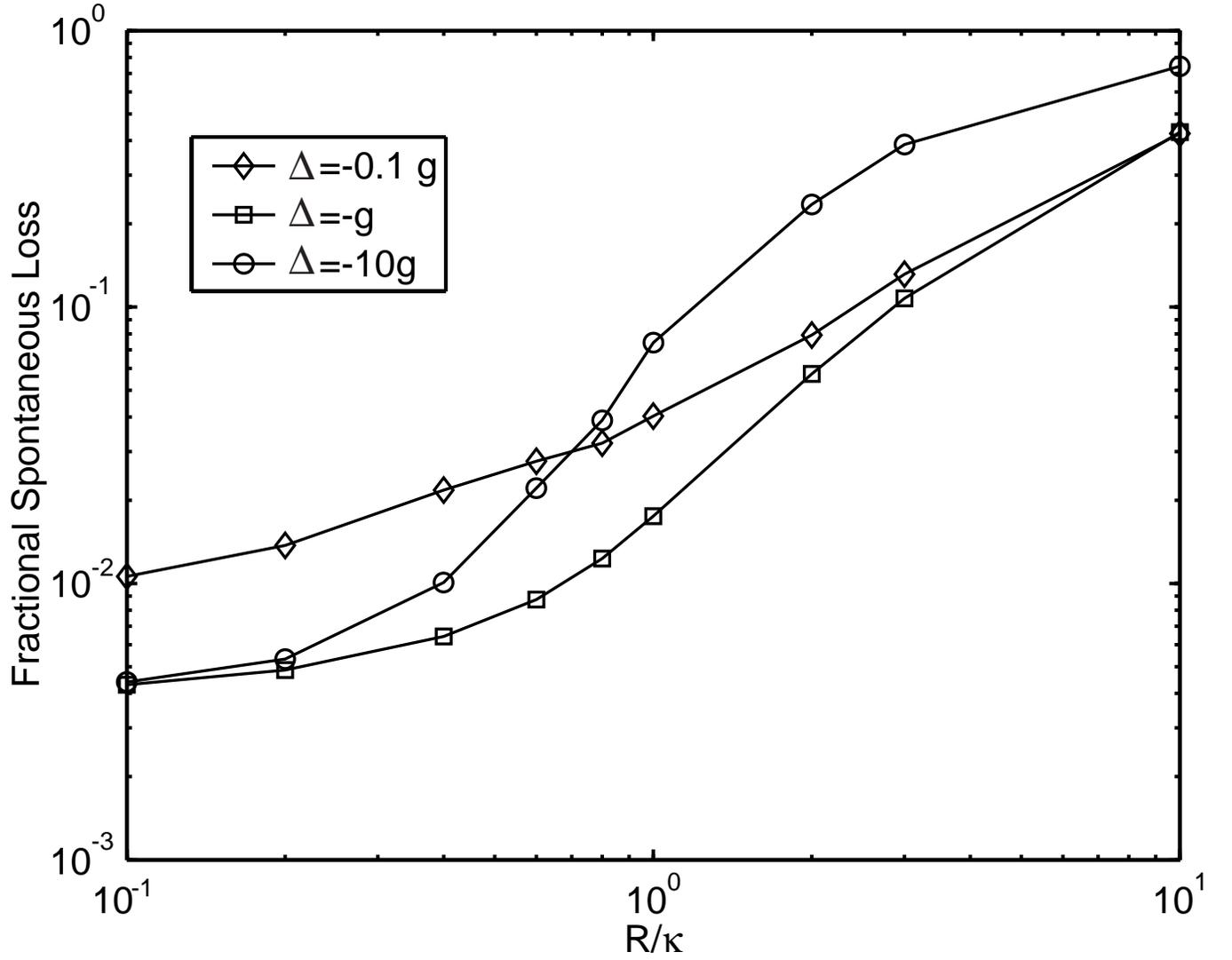}
\caption{Fractional spontaneous loss  as a function of ramp rate, $R$, and 
detuning $\Delta$ ($\gamma=g/20,$ and $\kappa=g/10$).
  As expected, larger ramp rates lead to an increased loss when the system can 
no longer adiabatically follow the dark state. Note how the dependence on 
$\Delta$ varies with $R$. In the adiabatic limit, small $R$, large $\Delta$ 
yields the lowest fractional spontaneous loss. However, when one ramps the 
system quickly, {\it i.e.}, at large $R$, the dependence on $\Delta$ dependence 
is reversed and small $\Delta$ yields the lowest loss.}
\label{fig:rerr}
\end{figure}
\begin{figure}
\includegraphics{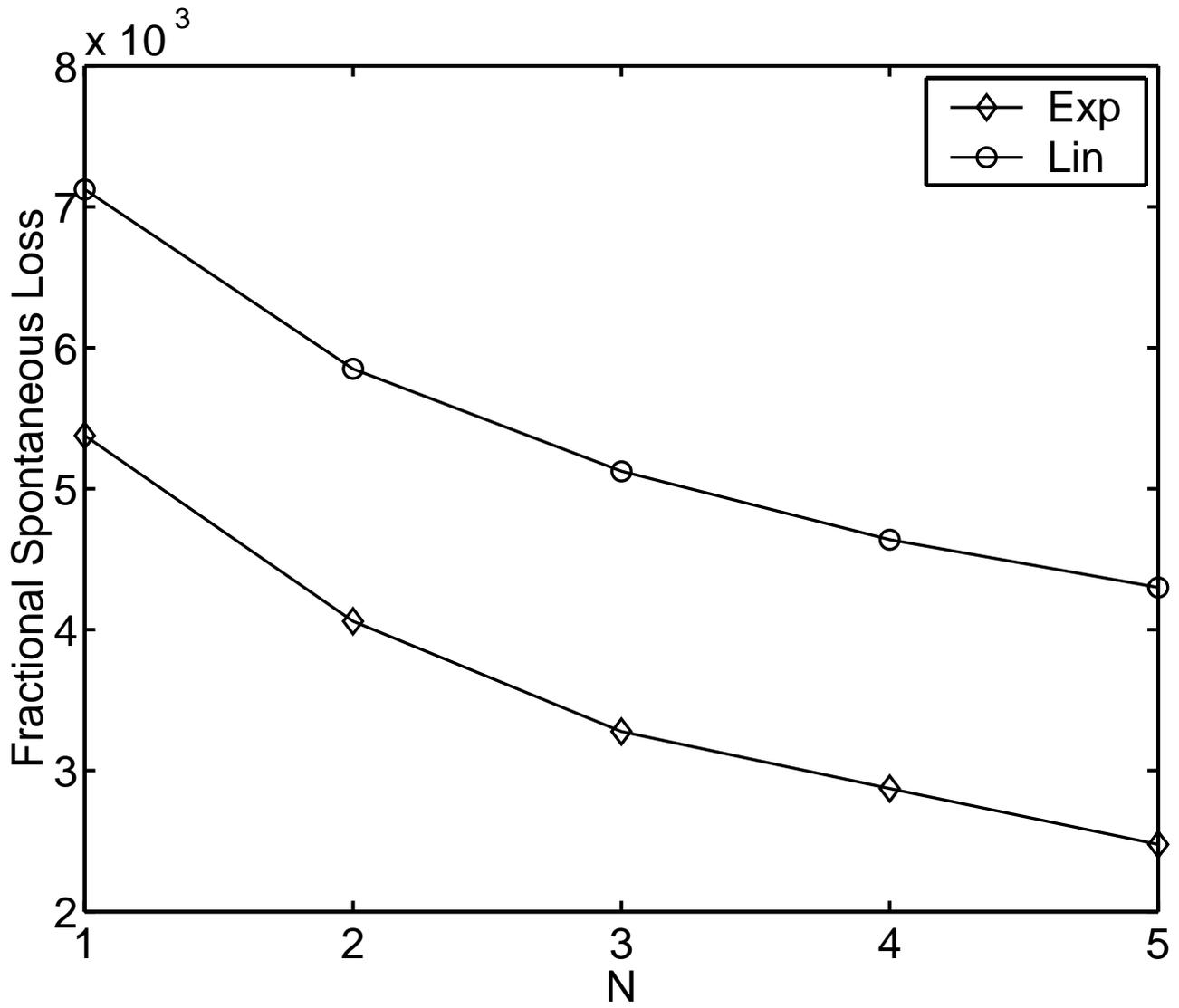}
 n\caption{Fractional spontaneous loss for linear and Gaussian ramping 
($\Delta=-2g,\gamma=g/20,$ and $\kappa=g/10$). For the linear ramp (circles), 
$R=g/30$. For the Gaussian ramp (diamonds), $r(t)=g\exp{(t-t_0)^2/(2\tau^2)}$, 
$\tau=50/g$. The Gaussian width, $\tau$, was chosen such that the process of 
emitting a single photon would occur in the same time as the linear ramp.  One 
expects that a more sophisticated pulse could result in an order of magnitude 
reduction of the fractional spontaneous loss.}
\label{fig:nexp}
\end{figure} 
\begin{figure}
\includegraphics{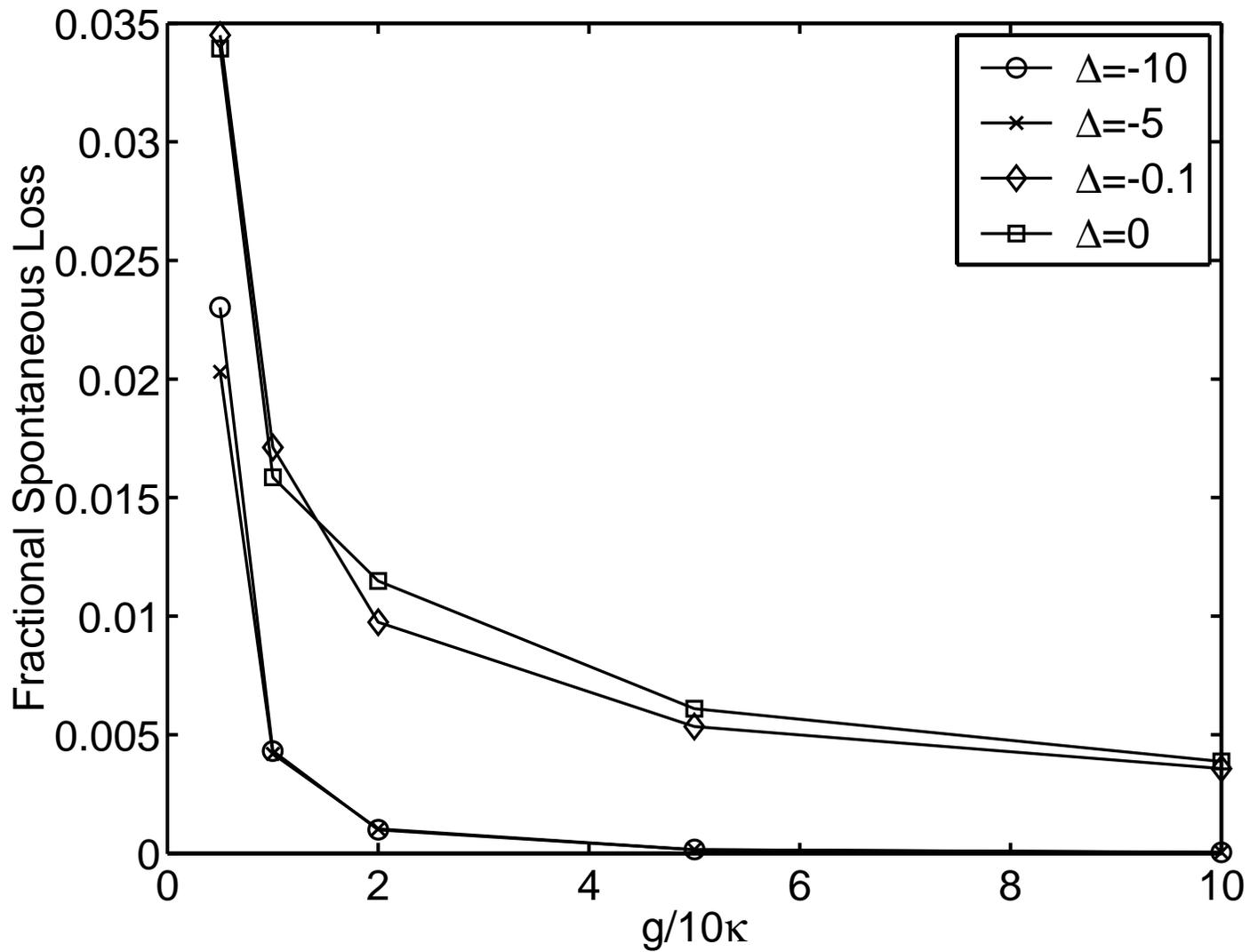}
\caption{Fractional spontaneous loss as a function of cavity coupling, $g$, and 
detuning, $\Delta$ ($R=g/100,\gamma=\kappa/2$).  The values of $\Delta$ in the 
legend are in units of $10 \kappa$.  One sees that at all values of $\Delta$ 
there is a significant decrease in fractional spontaneous loss for increasing 
$g$.  In the large $|\Delta|$ limit, one finds that the decrease in spontaneous 
loss with $g$ has a $1/g^2$ dependence, as predicted by our pertubative analysis 
(see Section \ref{sec:small}).}

\label{fig:gerr}
\end{figure}


\begin{thebibliography}{22}
\expandafter\ifx\csname natexlab\endcsname\relax\def\natexlab#1{#1}\fi
\expandafter\ifx\csname bibnamefont\endcsname\relax
  \def\bibnamefont#1{#1}\fi
\expandafter\ifx\csname bibfnamefont\endcsname\relax
  \def\bibfnamefont#1{#1}\fi
\expandafter\ifx\csname citenamefont\endcsname\relax
  \def\citenamefont#1{#1}\fi
\expandafter\ifx\csname url\endcsname\relax
  \def\url#1{\texttt{#1}}\fi
\expandafter\ifx\csname urlprefix\endcsname\relax\def\urlprefix{URL }\fi

\providecommand{\bibinfo}[2]{#2}
\providecommand{\eprint}[2][]{\url{#2}}

\bibitem[{\citenamefont{Gisin et~al.}(2002)\citenamefont{Gisin, Ribordy, Tittel, 
and Zbinden}}]{qcryptrev}
\bibinfo{author}{\bibfnamefont{N.}~\bibnamefont{Gisin}},
  \bibinfo{author}{\bibfnamefont{G.}~\bibnamefont{Ribordy}},
  \bibinfo{author}{\bibfnamefont{W.}~\bibnamefont{Tittel}}, \bibnamefont{and}
  \bibinfo{author}{\bibfnamefont{H.}~\bibnamefont{Zbinden}},
  \bibinfo{journal}{Rev. Mod. Phys. } \textbf{\bibinfo{volume}{74}},
  \bibinfo{pages}{145} (\bibinfo{year}{2002}).

\bibitem[{\citenamefont{Parkins et~al.}(2001)\citenamefont{Parkins, Marte, 
Zoller,
and Kimble }}]{parkins93}
\bibinfo{author}{\bibfnamefont{A. S.}~\bibnamefont{Parkins}},
\bibinfo{author}{\bibfnamefont{P.}~\bibnamefont{Marte}}, 
\bibinfo{author}{\bibfnamefont{P.}~\bibnamefont{Zoller}}, 
\bibnamefont{and}
  \bibinfo{author}{\bibfnamefont{H. J.}~\bibnamefont{Kimble}},
  \bibinfo{journal}{Phys. Rev. Lett.} \textbf{\bibinfo{volume}{71}}, 
\bibinfo{pages}{3095}
  (\bibinfo{year}{1993}).

\bibitem[{\citenamefont{Knill et~al.}(2001)\citenamefont{Knill, Laflamme, and
  Milburn}}]{knil01linear}
\bibinfo{author}{\bibfnamefont{E.}~\bibnamefont{Knill}},
  \bibinfo{author}{\bibfnamefont{R.}~\bibnamefont{Laflamme}}, \bibnamefont{and}
  \bibinfo{author}{\bibfnamefont{G.}~\bibnamefont{Milburn}},
  \bibinfo{journal}{Nature} \textbf{\bibinfo{volume}{409}}, \bibinfo{pages}{46}
  (\bibinfo{year}{2001}).

\bibitem[{\citenamefont{Kim et~al.}(1999)\citenamefont{Kim, Benson, Kan, and
  Yamamoto}}]{kim99single}
\bibinfo{author}{\bibfnamefont{J.}~\bibnamefont{Kim}},
  \bibinfo{author}{\bibfnamefont{O.}~\bibnamefont{Benson}},
  \bibinfo{author}{\bibfnamefont{H.}~\bibnamefont{Kan}}, \bibnamefont{and}
  \bibinfo{author}{\bibfnamefont{Y.}~\bibnamefont{Yamamoto}},
  \bibinfo{journal}{Nature} \textbf{\bibinfo{volume}{397}},
  \bibinfo{pages}{500} (\bibinfo{year}{1999}).

\bibitem[{\citenamefont{Michler et~al.}(2000)\citenamefont{Michler, Kiraz,
  Becher, Schoenfeld, Petroff, Zhang, Hu, and Imamoglu}}]{mich00single}
\bibinfo{author}{\bibfnamefont{P.}~\bibnamefont{Michler}},
  \bibinfo{author}{\bibfnamefont{A.}~\bibnamefont{Kiraz}},
  \bibinfo{author}{\bibfnamefont{C.}~\bibnamefont{Becher}},
  \bibinfo{author}{\bibfnamefont{W.~V.} \bibnamefont{Schoenfeld}},
  \bibinfo{author}{\bibfnamefont{P.~M.} \bibnamefont{Petroff}},
  \bibinfo{author}{\bibfnamefont{L.}~\bibnamefont{Zhang}},
  \bibinfo{author}{\bibfnamefont{E.}~\bibnamefont{Hu}}, \bibnamefont{and}
  \bibinfo{author}{\bibfnamefont{A.}~\bibnamefont{Imamoglu}},
  \bibinfo{journal}{Science} \textbf{\bibinfo{volume}{290}},
  \bibinfo{pages}{2282} (\bibinfo{year}{2000}).

\bibitem[{\citenamefont{Yuan et~al.}(2001)\citenamefont{Yuan, Kardynal,
  Stevenson, Shields, Lobo, Cooper, Beattie, Ritchie, and Pepper}}]{yuan01}
\bibinfo{author}{\bibfnamefont{Z.}~\bibnamefont{Yuan}},
  \bibinfo{author}{\bibfnamefont{B.~E.} \bibnamefont{Kardynal}},
  \bibinfo{author}{\bibfnamefont{R.~M.} \bibnamefont{Stevenson}},
  \bibinfo{author}{\bibfnamefont{A.~J.} \bibnamefont{Shields}},

  \bibinfo{author}{\bibfnamefont{C.~J.} \bibnamefont{Lobo}},
  \bibinfo{author}{\bibfnamefont{K.}~\bibnamefont{Cooper}},
  \bibinfo{author}{\bibfnamefont{N.~S.} \bibnamefont{Beattie}},
  \bibinfo{author}{\bibfnamefont{D.~A.} \bibnamefont{Ritchie}},
  \bibnamefont{and} \bibinfo{author}{\bibfnamefont{M.}~\bibnamefont{Pepper}},
  \bibinfo{journal}{Science} \textbf{\bibinfo{volume}{295}},
  \bibinfo{pages}{102} (\bibinfo{year}{2001}).

\bibitem[{\citenamefont{Brunel et~al.}(1999)\citenamefont{Brunel, Lounis,
  Tamarat, and Orrit}}]{brun99single}
\bibinfo{author}{\bibfnamefont{C.}~\bibnamefont{Brunel}},
  \bibinfo{author}{\bibfnamefont{B.}~\bibnamefont{Lounis}},
  \bibinfo{author}{\bibfnamefont{P.}~\bibnamefont{Tamarat}}, \bibnamefont{and}
  \bibinfo{author}{\bibfnamefont{M.}~\bibnamefont{Orrit}},
  \bibinfo{journal}{Phys. Rev. Lett.} \textbf{\bibinfo{volume}{83}},
  \bibinfo{pages}{2722} (\bibinfo{year}{1999}).

\bibitem[{\citenamefont{Lounis and Moerner}(2000)}]{loun00single}
\bibinfo{author}{\bibfnamefont{B.}~\bibnamefont{Lounis}} \bibnamefont{and}
  \bibinfo{author}{\bibfnamefont{W.}~\bibnamefont{Moerner}},
  \bibinfo{journal}{Nature} \textbf{\bibinfo{volume}{407}},
  \bibinfo{pages}{491} (\bibinfo{year}{2000}).

\bibitem[{\citenamefont{Law and Kimble}(1997)}]{law97single}
\bibinfo{author}{\bibfnamefont{C.}~\bibnamefont{Law}} \bibnamefont{and}
  \bibinfo{author}{\bibfnamefont{H.}~\bibnamefont{Kimble}},
  \bibinfo{journal}{Journal of Modern Optics} \textbf{\bibinfo{volume}{44}},
  \bibinfo{pages}{2067} (\bibinfo{year}{1997}).

\bibitem[{\citenamefont{Kuhn et~al.}(1997)\citenamefont{Kuhn, Hennrich, Bondo,
  and Rempe}}]{kuhn99single}
\bibinfo{author}{\bibfnamefont{A.}~\bibnamefont{Kuhn}},
  \bibinfo{author}{\bibfnamefont{M.}~\bibnamefont{Hennrich}},
  \bibinfo{author}{\bibfnamefont{T.}~\bibnamefont{Bondo}}, \bibnamefont{and}
  \bibinfo{author}{\bibfnamefont{G.}~\bibnamefont{Rempe}},
  \bibinfo{journal}{App. Phys. B} \textbf{\bibinfo{volume}{B69}},
  \bibinfo{pages}{373} (\bibinfo{year}{1999}).

\bibitem[{\citenamefont{Hennrich et~al.}(2000)\citenamefont{Hennrich, Legero,
  Kuhn, and Rempe}}]{henn00raman}
\bibinfo{author}{\bibfnamefont{M.}~\bibnamefont{Hennrich}},
  \bibinfo{author}{\bibfnamefont{T.}~\bibnamefont{Legero}},
  \bibinfo{author}{\bibfnamefont{A.}~\bibnamefont{Kuhn}}, \bibnamefont{and}
  \bibinfo{author}{\bibfnamefont{G.}~\bibnamefont{Rempe}},
  \bibinfo{journal}{Phys. Rev. Lett.} \textbf{\bibinfo{volume}{85}},
  \bibinfo{pages}{4872} (\bibinfo{year}{2000}).

\bibitem[{\citenamefont{Kuhn et~al.}(2002)\citenamefont{Kuhn, Hennrich,
  and Rempe}}]{rempexp}
\bibinfo{author}{\bibfnamefont{A.}~\bibnamefont{Kuhn}},
  \bibinfo{author}{\bibfnamefont{M.}~\bibnamefont{Hennrich}},
\bibnamefont{and}
  \bibinfo{author}{\bibfnamefont{G.}~\bibnamefont{Rempe}},
  \bibinfo{journal}{Phys. Rev. Lett.} \textbf{\bibinfo{volume}{89}},
  \bibinfo{pages}{067901} (\bibinfo{year}{2002}).

\bibitem[{\citenamefont{Lvovsky et~al.}(2001)\citenamefont{Lvovsky, Hansen,
  Aichele, Benson, Mlynek, and Schiller}}]{lvov01fock}
\bibinfo{author}{\bibfnamefont{A.I.}~\bibnamefont{Lvovsky}},
  \bibinfo{author}{\bibfnamefont{H.}~\bibnamefont{Hansen}},
  \bibinfo{author}{\bibfnamefont{T.}~\bibnamefont{Aichele}},
  \bibinfo{author}{\bibfnamefont{O.}~\bibnamefont{Benson}},
  \bibinfo{author}{\bibfnamefont{J.}~\bibnamefont{Mlynek}}, \bibnamefont{and}
  \bibinfo{author}{\bibfnamefont{S.}~\bibnamefont{Schiller}},
  \bibinfo{journal}{Phys. Rev. Lett.} \textbf{\bibinfo{volume}{87}},
  \bibinfo{pages}{050402} (\bibinfo{year}{2001}).

\bibitem[{\citenamefont{Wineland et~al.}(1992)\citenamefont{Wineland,
  Bollinger, Itano, Moore, and Heinzen}}]{wine92squeeze}
\bibinfo{author}{\bibfnamefont{D.J.}~\bibnamefont{Wineland}},
  \bibinfo{author}{\bibfnamefont{J.J.}~\bibnamefont{Bollinger}},
  \bibinfo{author}{\bibfnamefont{W.M.}~\bibnamefont{Itano}},
  \bibinfo{author}{\bibfnamefont{F.L.}~\bibnamefont{Moore}}, \bibnamefont{and}
  \bibinfo{author}{\bibfnamefont{D.J.}~\bibnamefont{Heinzen}},
  \bibinfo{journal}{Phys. Rev. A} \textbf{\bibinfo{volume}{46}},
  \bibinfo{pages}{R6797} (\bibinfo{year}{1992}).

\bibitem[{\citenamefont{Bouyer and Kasevich}(1997)}]{bouy97}
\bibinfo{author}{\bibfnamefont{P.}~\bibnamefont{Bouyer}} \bibnamefont{and}
  \bibinfo{author}{\bibfnamefont{M.A.}~\bibnamefont{Kasevich}},
  \bibinfo{journal}{Phys. Rev. A} \textbf{\bibinfo{volume}{56}},
  \bibinfo{pages}{R1083} (\bibinfo{year}{1997}).

\bibitem[{\citenamefont{Law and Eberly}(1996)}]{law96arb}
\bibinfo{author}{\bibfnamefont{C.K.}~\bibnamefont{Law}} \bibnamefont{and}
  \bibinfo{author}{\bibfnamefont{J.H.}~\bibnamefont{Eberly}},
  \bibinfo{journal}{Phys. Rev. Lett.} \textbf{\bibinfo{volume}{76}},
  \bibinfo{pages}{1055} (\bibinfo{year}{1996}).

\bibitem[{\citenamefont{Varcoe et~al.}(2000)\citenamefont{Varcoe, Brattke, and
  Walther}}]{varc00}
\bibinfo{author}{\bibfnamefont{B.}~\bibnamefont{Varcoe}},

  \bibinfo{author}{\bibfnamefont{S.}~\bibnamefont{Brattke}}, \bibnamefont{and}
  \bibinfo{author}{\bibfnamefont{H.}~\bibnamefont{Walther}},
  \bibinfo{journal}{Journal of Optics B: Quantum and Semiclassical Optics}
  \textbf{\bibinfo{volume}{2}}, \bibinfo{pages}{154} (\bibinfo{year}{2000}).

\bibitem[{\citenamefont{Domokos et~al.}(1998)\citenamefont{Domokos, Brune,
  Raimond, and Haroche}}]{domo98fock}
\bibinfo{author}{\bibfnamefont{P.}~\bibnamefont{Domokos}},
  \bibinfo{author}{\bibfnamefont{M.}~\bibnamefont{Brune}},
  \bibinfo{author}{\bibfnamefont{J.}~\bibnamefont{Raimond}}, \bibnamefont{and}
  \bibinfo{author}{\bibfnamefont{S.}~\bibnamefont{Haroche}},
  \bibinfo{journal}{European Physics Journal D} \textbf{\bibinfo{volume}{1}},
  \bibinfo{pages}{1} (\bibinfo{year}{1998}).

\bibitem[{\citenamefont{Bertet et~al.}(2002)\citenamefont{Bertet, Osnaghi,
  Milman, Auffeves, Maioli, Brune, Raimond, and Haroche}}]{bert02two}
\bibinfo{author}{\bibfnamefont{P.}~\bibnamefont{Bertet}},
  \bibinfo{author}{\bibfnamefont{S.}~\bibnamefont{Osnaghi}},
  \bibinfo{author}{\bibfnamefont{P.}~\bibnamefont{Milman}},
  \bibinfo{author}{\bibfnamefont{A.}~\bibnamefont{Auffeves}},
  \bibinfo{author}{\bibfnamefont{P.}~\bibnamefont{Maioli}},
  \bibinfo{author}{\bibfnamefont{M.}~\bibnamefont{Brune}},
  \bibinfo{author}{\bibfnamefont{J.~M.} \bibnamefont{Raimond}},
  \bibnamefont{and} \bibinfo{author}{\bibfnamefont{S.}~\bibnamefont{Haroche}},
  \bibinfo{journal}{Phys. Rev. Lett.} \textbf{\bibinfo{volume}{88}},
  \bibinfo{pages}{143601} (\bibinfo{year}{2002}).

\bibitem[{\citenamefont{Hood et~al.}(1998)\citenamefont{Hood, Chapman, Lynn,
  and Kimble}}]{hood98}
\bibinfo{author}{\bibfnamefont{C.J.}~\bibnamefont{Hood}},
  \bibinfo{author}{\bibfnamefont{M.S.}~\bibnamefont{Chapman}},
  \bibinfo{author}{\bibfnamefont{T.W.}~\bibnamefont{Lynn}}, \bibnamefont{and}
  \bibinfo{author}{\bibfnamefont{H.J.}~\bibnamefont{Kimble}},
  \bibinfo{journal}{Phys. Rev. Lett.} \textbf{\bibinfo{volume}{80}},
  \bibinfo{pages}{4157} (\bibinfo{year}{1998}).

\bibitem[{\citenamefont{S\"{o}rensen et~al.}(2001)\citenamefont{S\"{o}rensen,
  Duan, Cirac, and Zoller}}]{sore01ent}
\bibinfo{author}{\bibfnamefont{A.}~\bibnamefont{S\"{o}rensen}},
  \bibinfo{author}{\bibfnamefont{L.-M.} \bibnamefont{Duan}},
  \bibinfo{author}{\bibfnamefont{J.}~\bibnamefont{Cirac}}, \bibnamefont{and}
  \bibinfo{author}{\bibfnamefont{P.}~\bibnamefont{Zoller}},
  \bibinfo{journal}{Nature} \textbf{\bibinfo{volume}{409}}, \bibinfo{pages}{63}
  (\bibinfo{year}{2001}).

\bibitem[{\citenamefont{Raghavan et~al.}(2001)\citenamefont{Raghavan, Pu,
  Meystre, and Bigelow}}]{ragh01dicke}
\bibinfo{author}{\bibfnamefont{S.}~\bibnamefont{Raghavan}},
  \bibinfo{author}{\bibfnamefont{H.}~\bibnamefont{Pu}},
  \bibinfo{author}{\bibfnamefont{P.}~\bibnamefont{Meystre}}, \bibnamefont{and}
  \bibinfo{author}{\bibfnamefont{N.P.}~\bibnamefont{Bigelow}},
  \bibinfo{journal}{Optics Communications} \textbf{\bibinfo{volume}{188}},
  \bibinfo{pages}{249} (\bibinfo{year}{2001}).

\bibitem[{\citenamefont{Kuzmich et~al.}(2000)\citenamefont{Kuzmich, Mandel, and
  Bigelow}}]{kuzm00}
\bibinfo{author}{\bibfnamefont{A.}~\bibnamefont{Kuzmich}},
  \bibinfo{author}{\bibfnamefont{L.}~\bibnamefont{Mandel}}, \bibnamefont{and}
  \bibinfo{author}{\bibfnamefont{N.P.}~\bibnamefont{Bigelow}},
  \bibinfo{journal}{Phys. Rev. Lett.} \textbf{\bibinfo{volume}{85}},
  \bibinfo{pages}{1594} (\bibinfo{year}{2000}).

\bibitem[{\citenamefont{Pellizzari et~al.}(1995)\citenamefont{Pellizzari,
  Gardiner, Cirac, and Zoller}}]{peli95cqed}
\bibinfo{author}{\bibfnamefont{T.}~\bibnamefont{Pellizzari}},
  \bibinfo{author}{\bibfnamefont{S.A.}~\bibnamefont{Gardiner}},
  \bibinfo{author}{\bibfnamefont{J.I.}~\bibnamefont{Cirac}}, \bibnamefont{and}
  \bibinfo{author}{\bibfnamefont{P.}~\bibnamefont{Zoller}},
  \bibinfo{journal}{Phys. Rev. Lett.} \textbf{\bibinfo{volume}{75}},
  \bibinfo{pages}{3788} (\bibinfo{year}{1995}).
\bibitem[{g()}]{gfootnote}
\bibinfo{note}{We note that when identical values of g are not possible for all 
$N$ atoms one can still reliably produce N-photon states using a pump laser 
which does not distinguish between the atoms.  The analysis would be similar and 
rely on the dark states discussed in Ref. \cite{peli95cqed}}

\bibitem[{kap()}]{kappafootnote}
\bibinfo{note}{We assume here that photons are lost from the cavity by
  transmission through the cavity mirrors to traveling modes outside the
  cavity, thereby neglecting their possible absorption in the mirrors.}

\bibitem[{\citenamefont{Tavis et~al.}(1995)\citenamefont{Tavis and 
Cummings}}]{tavis}
\bibinfo{author}{\bibfnamefont{M.}~\bibnamefont{Tavis}} \bibnamefont{and}
  \bibinfo{author}{\bibfnamefont{F.W.}~\bibnamefont{Cummings}},
  \bibinfo{journal}{Phys. Rev. } \textbf{\bibinfo{volume}{170}},
  \bibinfo{pages}{379} (\bibinfo{year}{1968}).

\bibitem[{\citenamefont{Carmichael}(1993)}]{carmichael}
\bibinfo{author}{\bibfnamefont{H.J.}~\bibnamefont{Carmichael}},
  \emph{\bibinfo{title}{An open system approach to quantum optics}}
  (\bibinfo{publisher}{Springer, Berlin, UK},
  \bibinfo{year}{2000}).

\bibitem[{\citenamefont{Plenio and Knight}(1995)}]{plenioqj}
\bibinfo{author}{\bibfnamefont{M.B.}~\bibnamefont{Plenio}} \bibnamefont{and}
  \bibinfo{author}{\bibfnamefont{P.L.}~\bibnamefont{Gardiner}},
  \bibinfo{journal}{Rev. Mod. Phys. } \textbf{\bibinfo{volume}{70}},
  \bibinfo{pages}{101} (\bibinfo{year}{1998}).

\bibitem[{\citenamefont{Wiseman and Milburn}(1993)}]{wiseman93}
\bibinfo{author}{\bibfnamefont{H.M.}~\bibnamefont{Wiseman}} \bibnamefont{and}
  \bibinfo{author}{\bibfnamefont{G.J.}~\bibnamefont{Milburn}},
  \bibinfo{journal}{Phys. Rev. A.} \textbf{\bibinfo{volume}{47}},
  \bibinfo{pages}{1652} (\bibinfo{year}{1993}).

\bibitem[{\citenamefont{Beidenharn and Van Dam}(1965)}]{schwinger}
\bibinfo{editor}{\bibfnamefont{L.C.}~\bibnamefont{Biedenharn}} \bibnamefont{and}
  \bibinfo{editor}{\bibfnamefont{H.}~\bibnamefont{Van Dam}},
  \emph{\bibinfo{title}{Quantum Theory of Angular Momentum}}
  (\bibinfo{publisher}{Academic Press, New York},
  \bibinfo{year}{1965}).

\bibitem[{\citenamefont{Sakurai}(1994)}]{sakurai}
\bibinfo{author}{\bibfnamefont{J.J.}~\bibnamefont{Sakurai}},
  \emph{\bibinfo{title}{Modern Quantum Mechanics}}
  (\bibinfo{publisher}{Addison-Wesley, New York},
  \bibinfo{year}{1994}).


\bibitem[{\citenamefont{Messiah}(1961)}]{messiahref}
\bibinfo{author}{\bibfnamefont{A.}~\bibnamefont{Messiah}},
  \emph{\bibinfo{title}{Quantum Mechanics}}
  (\bibinfo{publisher}{Interscience, New York},
  \bibinfo{year}{1961}).

\bibitem[{\citenamefont{Ye, Vernooy, and Kimble}(1999)}]{kimcavpar}
\bibinfo{author}{\bibfnamefont{J.}~\bibnamefont{Ye}},
  \bibinfo{author}{\bibfnamefont{D.W.}~\bibnamefont{Vernooy}}, \bibnamefont{and}
  \bibinfo{author}{\bibfnamefont{H.J.}~\bibnamefont{Kimble}}, 
 \bibinfo{journal}{Phys. Rev. Lett.} \textbf{\bibinfo{volume}{83}},
  \bibinfo{pages}{4987} (\bibinfo{year}{1999}).

\bibitem[{\citenamefont{Kimble}(1998)}]{Kimble:98}
\bibinfo{author}{\bibfnamefont{H.J.}~\bibnamefont{Kimble}},
  \bibinfo{journal}{Physica Scripta } \textbf{\bibinfo{volume}{T76}},
  \bibinfo{pages}{127} (\bibinfo{year}{1998}).

\bibitem[{\citenamefont{Holland and Burnett}(1961)}]{hlimit}
\bibinfo{author}{\bibfnamefont{M.J.}~\bibnamefont{Holland}},
 \bibnamefont{and}
  \bibinfo{author}{\bibfnamefont{K.}~\bibnamefont{Burnett}}, 
 \bibinfo{journal}{Phys. Rev. Lett.} \textbf{\bibinfo{volume}{71}},
  \bibinfo{pages}{1355} (\bibinfo{year}{1993}).
\bibitem[{\citenamefont{Kok, Lee, and Dowling}(1961)}]{Dowling:02}
\bibinfo{author}{\bibfnamefont{P.}~\bibnamefont{Kok}},
  \bibinfo{author}{\bibfnamefont{H.}~\bibnamefont{Lee}}, \bibnamefont{and}
  \bibinfo{author}{\bibfnamefont{J.P.}~\bibnamefont{Dowling}}, 
 \bibinfo{journal}{Phys. Rev. A.} \textbf{\bibinfo{volume}{65}},
  \bibinfo{pages}{052104} (\bibinfo{year}{2002}).
\bibitem[{\citenamefont{Bollinger et. al.}(1996)}]{freqmet}
\bibinfo{author}{\bibfnamefont{J.J.}~\bibnamefont{Bollinger}}, 
\bibinfo{author}{\bibnamefont{W.M.}~\bibnamefont{Itano}}, 
\bibinfo{author}{\bibnamefont{D.J.}~\bibnamefont{Wineland}},  \bibnamefont{and}
  \bibinfo{author}{\bibfnamefont{D.J.}~\bibnamefont{Heinzen}},
  \bibinfo{journal}{Phys. Rev. A} \textbf{\bibinfo{volume}{54}},
  \bibinfo{pages}{R4649} (\bibinfo{year}{1996}).

\bibitem[{\citenamefont{Dowling}(1998)}]{qgyro}
\bibinfo{author}{\bibfnamefont{J.P.}~\bibnamefont{Dowling}},
  \bibinfo{journal}{Phys. Rev. A} \textbf{\bibinfo{volume}{57}}  
\bibinfo{pages}{4736} (\bibinfo{year}{1998}).

\bibitem[{\citenamefont{Jozsa et. al.}(2000)}]{clocksync}
\bibinfo{author}{\bibfnamefont{R.}~\bibnamefont{Jozsa}}, 
\bibinfo{author}{\bibnamefont{D.S.}~\bibnamefont{Abrams}}, 
\bibinfo{author}{\bibnamefont{J.P.}~\bibnamefont{Dowling}},  \bibnamefont{and}
  \bibinfo{author}{\bibfnamefont{C.P.}~\bibnamefont{Williams}},
  \bibinfo{journal}{Phys. Rev. Lett.} \textbf{\bibinfo{volume}{85}},
  \bibinfo{pages}{2010} (\bibinfo{year}{2000}).

\bibitem[{\citenamefont{Rice and Zhao}(1994)}]{rice}
\bibinfo{author}{\bibfnamefont{S.A.}~\bibnamefont{Rice}}, \bibnamefont{and}
  \bibinfo{author}{\bibfnamefont{M.}~\bibnamefont{Zhao}},
  \emph{\bibinfo{title}{Optical Control of Molecular Dynamics}}
  (\bibinfo{publisher}{Wiley, New York},
  \bibinfo{year}{2000}).

\bibitem[{\citenamefont{Dessent and Muller-Dethfels}(2000)}]{REMPI}
\bibinfo{author}{\bibfnamefont{C.E.H.}~\bibnamefont{Dessent}},
 \bibnamefont{and}
  \bibinfo{author}{\bibfnamefont{K.}~\bibnamefont{Muller-Dethlefs}}, 
 \bibinfo{journal}{Chem. Rev.} \textbf{\bibinfo{volume}{100}},
  \bibinfo{pages}{3999} (\bibinfo{year}{2000}).

\bibitem[{\citenamefont{Lukin et~al.}(2000)\citenamefont{Lukin, Yelin, and 
Fleischauer}}]{lukin}
\bibinfo{author}{\bibfnamefont{M.D.}~\bibnamefont{Lukin}},
  \bibinfo{author}{\bibfnamefont{S.F.}~\bibnamefont{Yelin}},
\bibnamefont{and}
  \bibinfo{author}{\bibfnamefont{M.}~\bibnamefont{Fleischhauer}},
  \bibinfo{journal}{Phys. Rev. Lett.} \textbf{\bibinfo{volume}{84}},
  \bibinfo{pages}{4232} (\bibinfo{year}{2000}).

\end{thebibliography}
\end{document}